\RequirePackage{fix-cm}
\documentclass[smallextended,envcountsect]{svjour3} %
\smartqed  %
\usepackage{amsmath,amssymb,amsfonts}
\usepackage{algorithmic}
\usepackage{textcomp}
\usepackage{xcolor}
\usepackage{graphicx}
\usepackage{multirow}
\usepackage{makecell}
\usepackage{csquotes}
\usepackage{framed}
\usepackage{array}
\usepackage{textcomp}
\usepackage[bookmarks=false]{hyperref}
\usepackage[capitalise]{cleveref}
\usepackage{dblfloatfix}    %
\usepackage{natbib}
\usepackage{soul}
\definecolor{greenncs}{rgb}{0.0, 0.62, 0.42}
\definecolor{airforceblue}{rgb}{0.36, 0.54, 0.66}
\colorlet{soulc}{greenncs!30}
\sethlcolor{soulc}

\newcolumntype{L}[1]{>{\raggedright\let\newline\\\arraybackslash\hspace{0pt}}m{#1}}
\newcolumntype{C}[1]{>{\centering\let\newline\\\arraybackslash\hspace{0pt}}m{#1}}
\newcolumntype{R}[1]{>{\raggedleft\let\newline\\\arraybackslash\hspace{0pt}}m{#1}}

\crefformat{footnote}{#2\footnotemark[#1]#3}
\begin{document}

\title{Automatic Identification of Self-Admitted\\
Technical Debt from Four Different Sources
}

\author{Yikun Li         \and
        Mohamed Soliman  \and
        Paris Avgeriou
}

\institute{Yikun Li \at
              Bernoulli Institute for Mathematics, Computer Science and Artificial Intelligence, University of Groningen, The Netherlands \\
              \email{yikun.li@rug.nl}
           \and
           Mohamed Soliman \at
              Bernoulli Institute for Mathematics, Computer Science and Artificial Intelligence, University of Groningen, The Netherlands \\
              \email{m.a.m.soliman@rug.nl}
           \and
           Paris Avgeriou \at
              Bernoulli Institute for Mathematics, Computer Science and Artificial Intelligence, University of Groningen, The Netherlands \\
              \email{p.avgeriou@rug.nl}
           \and
}

\date{Received: date / Accepted: date}

\maketitle

\begin{abstract}
Technical debt refers to taking shortcuts to achieve short-term goals while sacrificing the long-term maintainability and evolvability of software systems.
A large part of technical debt is explicitly reported by the developers themselves; this is commonly referred to as \emph{Self-Admitted Technical Debt} or \emph{SATD}. 
Previous work has focused on identifying SATD from source code comments and issue trackers. However, there are no approaches available for automatically identifying SATD from other sources such as commit messages and pull requests, or by combining multiple sources.
Therefore, we propose and evaluate an approach for automated SATD identification that integrates four sources: source code comments, commit messages, pull requests, and issue tracking systems.
Our findings show that our approach outperforms baseline approaches and achieves an average F1-score of 0.611 when detecting four types of SATD (i.e., code/design debt, requirement debt, documentation debt, and test debt) from the four aforementioned sources.
Thereafter, we analyze 23.6M code comments, 1.3M commit messages, 3.7M issue sections, and 1.7M pull request sections to characterize SATD in 103 open-source projects.
Furthermore, we investigate the SATD keywords and relations between SATD in different sources.
The findings indicate, among others, that: 1) SATD is evenly spread among all sources; 2) issues and pull requests are the two most similar sources regarding the number of shared SATD keywords, followed by commit messages, and then followed by code comments; 3) there are four kinds of relations between SATD items in the different sources.

\end{abstract}

\section{Introduction}
\label{sec:introduction}

Technical debt is a metaphor expressing the compromise of maintainability and evolvability of software systems in the long term, in order to achieve short-term goals \citep{avgeriou_et_al:DR:2016:6693}.
If technical debt is ignored and not proactively managed, it tends to accumulate, potentially resulting in a maintenance crisis \citep{allman2012managing}.
There are several activities involved in technical debt management, the first of which is its identification \citep{li2015systematic}: distinguishing those sub-optimal software artifacts that hinder maintenance and evolution activities.

Most of the previous studies on identifying technical debt have focused on static source code analysis \citep{alves2014towards,li2015systematic}.
While such approaches are effective in detecting technical debt at the code level, they are less so in identifying other types, such as documentation debt or requirement debt.
This was partially remedied, when Potdar and Shihab found out that developers frequently use code comments, such as \emph{TODO} or \emph{Fixme}, to indicate the existence of technical debt \citep{potdar2014exploratory}.
They called attention to this special kind of technical debt, known as \emph{Self-Admitted Technical Debt} or \emph{SATD}, as it is explicitly admitted by developers in software artifacts.
Making SATD explicit has been shown to be an important and valuable complement to static code analysis, especially for detecting technical debt types other than the code level \citep{SIERRA201970}.

The identification of SATD has been fairly well researched, with the vast majority of this work focusing on source code comments \citep{da2017using,huang2018identifying,ren2019neural,wang2020detecting}; there are also a couple of studies that identify SATD from issue tracking systems \citep{dai2017detecting,li2022identifying}.
However, besides code comments and issue trackers, Zampetti et al. \citep{zampetti2021self} found that technical debt is commonly documented in other sources as well, such as commit messages and pull requests;  this holds for both industry and open-source projects. %
Nevertheless, there are no approaches for identifying SATD from commit messages and pull requests.
Furthermore, all previous approaches for SATD identification use only a single data source, i.e., either source code comments or issue trackers.
This paper attempts to address these shortcomings by proposing an integrated approach to automatically identify SATD from four different sources (i.e., source code comments, issue trackers, commit messages, and pull requests). We focus on these four sources as they are the four most popular software ones for self-admitting technical debt \citep{zampetti2021self}.

Using an integrated approach to detect SATD from different sources has three advantages over using multiple identifiers.
First, it would be simpler, more lightweight, and easier to use.
Researchers would train the integrated approach once to identify SATD from different sources instead of training multiple machine learning models for different sources.
As for practitioners, they would be able to use one identifier to detect SATD from distinct sources instead of multiple SATD identifiers.
Second, the integrated identifier would be more extensible.
In our previous study \citep{li2022identifying}, we discovered similarities between SATD in source code comments and issue tracking systems.
To incorporate a new source (e.g., mailing lists) into the integrated approach, the knowledge of SATD from current sources learned by the integrated approach could be used to improve the predictive performance of the new source.
Third, the integrated approach can learn to identify SATD from different sources in parallel while learning a shared representation, which could potentially improve the predictive performance of identifying SATD from individual sources.

The proposed SATD identification approach in this paper is trained and tested, and subsequently compared to three traditional machine learning approaches (i.e., logistic regression, support vector machine, and random forest), one deep learning approach (i.e., Text-CNN), and one baseline approach (i.e., random baseline).
The training requires datasets; while there are SATD datasets for source code comments and issues, there are no datasets available for commit messages and pull requests. %
We thus collect 5,000 commit messages and 5,000 pull request sections from 103 open-source projects from the Apache ecosystem.
Then we manually classify the collected data into different types of SATD or non-SATD according to the classification framework by \cite{li2022identifying}. 
After training and evaluating the classifier, we summarize and present lists of keywords for different types of SATD and SATD from different sources.
Next, we demonstrate the characteristics of SATD in 103 open-source projects.
Finally, we explore the relations between SATD in different sources.

The main contributions of this paper are described as follows:

\begin{itemize}
    \item \textbf{Contributing rich datasets.}
    We created a SATD dataset containing 5,000 commit messages and 5,000 pull request sections from 103 Apache open-source projects.
    We manually tagged each item in this dataset as non-SATD or SATD (including types of SATD).
    Moreover, we also created a large dataset containing 23.7M code comments, 1.3M commit messages, 0.3M pull requests, and 0.6M issues from the same 103 Apache open-source projects.
    We make these two datasets publicly available\footnote{\label{l:data}\url{https://github.com/yikun-li/satd-different-sources-data})} to facilitate research in this area.
    
    \item \textbf{Proposing an approach (MT-Text-CNN) to identify four types of SATD from four sources.}
    This approach is based on a convolutional neural network and leverages the multitask learning technique.
    The results indicate that our MT-Text-CNN approach achieves an average F1-score of 0.611 when identifying four types of SATD from the four aforementioned sources, outperforming other baseline methods by a large margin.
    
    \item \textbf{Summarizing lists of SATD keywords.}
    SATD keywords for different types of SATD and for SATD from different sources are presented.
    The numbers of shared keywords between different sources are also calculated.
    The results show that issues and pull requests are the two most similar sources concerning the number of shared keywords, followed by commit messages, and finally by code comments.
    
    \item \textbf{Characterizing SATD from different sources in 103 open-source projects.}
    The proposed MT-Text-CNN approach is utilized to identify SATD from 103 open-source projects.
    The number and percentage of different types of SATD are presented. 
    The results indicate that SATD is evenly spread among different sources.
    
    \item \textbf{Investigating relations between SATD in different sources.}
    We analyzed a sample of the identified SATD to explore the relations between SATD in different sources.
    The results show that there are four types of relations between SATD in different sources.
\end{itemize}

The remainder of this paper is organized as follows. 
In \cref{sec:related}, related work is presented.
\cref{sec:approach} elaborates on the study design, while the results are reported in \cref{sec:results}.
The study results are subsequently discussed in \cref{sec:discussion} and threats to validity are assessed in \cref{sec:validity}.
Finally, conclusions are drawn in \cref{sec:conclusion}.

\section{Related Work}
\label{sec:related}

In this work, we focus on automatically identifying SATD from different sources.
Thus, we explore related work from two areas: work associated with managing SATD in different sources and work associated with \emph{automatic} SATD identification.

\subsection{Self-Admitted Technical Debt in Different Sources}

Several studies have indicated that technical debt can be admitted by developers in different sources \citep{SIERRA201970,zampetti2021self}.
Zampetti \textit{et al.} \citep{zampetti2021self} surveyed 101 software developers to study the SATD practices in industrial and open-source projects.
The results showed that source code comments are the most popular sources for documenting SATD, followed by commit messages, pull requests, issue trackers, private documents, etc.

Among all sources, the majority of research has focused on SATD in source code comments \citep{SIERRA201970}.
Potdar and Shihab were the first to shed light on self-admitted technical debt in this source \citep{potdar2014exploratory}.
They analyzed source code comments of four large open-source software projects to identify SATD.
They found that 2.4\% to 31.0\% of analyzed files contain SATD and experienced developers tend to introduce more SATD compared to others.
Subsequently, \cite{maldonado2015detecting} examined code comments in five open-source projects to investigate the different types of SATD.
The results classify SATD into five types, namely design, defect, documentation, test, and requirement debt.
The most common type of SATD is design debt which ranges from 42\% to 84\% in different projects.
Furthermore, \cite{kamei2016using} analyzed the source code comments of JMeter and studied the interest of the SATD.
They found that 42\% to 44\% of the SATD generates positive interest (debt that needs more effort to be repaid in the future).

Apart from source code comments, issue tracking systems are the second most common source for studying SATD \citep{bellomo2016got,dai2017detecting,li2020identification,xavier2020beyond,li2022identifying}.
\cite{bellomo2016got} analyzed 1,264 issues from four issue tracking systems and found 109 SATD issues.
They found that issues could also contain SATD even if they are not tagged as technical debt issues.
Subsequently, in our previous work \citep{li2020identification}, we manually examined 500 issues from two open-source projects and found eight types of SATD from issue trackers, namely architecture, build, code, defect, design, documentation, requirement, and test debt.
The results indicated that developers report SATD in issues in three different points in time and most of SATD is repaid after introduction.
Additionally, \cite{xavier2020beyond} studied a sample of 286 issues from five open-source projects.
They found that 29\% of SATD issues can be traced back to source code comments, and SATD issues take more time to be closed compared to non-SATD issues.

Moreover, there are limited studies that make use of the information in commit messages to study SATD \citep{zampetti2018self,iammarino2019self,iammarino2021empirical}.
To investigate SATD repayment, a quantitative and qualitative study was conducted by \cite{zampetti2018self}. 
They explored to which extent SATD removal is documented in commit messages in five-open source projects.
They analyzed the textual similarity between the SATD code comments and corresponding commit messages to determine whether SATD removals are confirmed in commit messages.
The results revealed that about 8\% of SATD removals are documented in commit messages, while between 20\% and 50\% of SATD comments are removed by accident.
\cite{iammarino2019self,iammarino2021empirical} investigated the relationship between refactoring actions and SATD removal by analyzing four open-source projects.
The results indicated that refactoring operations are more likely to occur in conjunction with SATD removals than with other changes.

\subsection{Automatic Identification of Self-Admitted Technical Debt}

There are numerous studies that focus on automatically identifying SATD, the vast majority of which use source code comments \citep{da2017using,huang2018identifying,ren2019neural,wang2020detecting,chen2021multiclass}.
The study by \cite{da2017using} was the first to explore automatic SATD identification.
They trained two maximum entropy classifiers to detect design and requirement SATD from code comments and presented a list of keywords of SATD comments.
Subsequently, \cite{huang2018identifying} proposed a text-mining-based approach to classify SATD and non-SATD source code comments.
Specifically, they utilized feature selection and ensemble learning techniques to improve predictive performance.
Thereafter, \cite{ren2019neural} introduced a convolutional neural network-based approach to improving the identification performance, while \cite{wang2020detecting} explored the efficiency of an attention-based approach in SATD identification.
Additionally, \cite{chen2021multiclass} trained an XGBoost classifier to identify three types of SATD, namely design, defect, and requirement debt from source code comments.
It is noted that apart from the studies by \cite{da2017using} and \cite{chen2021multiclass} that detect different types of SATD, the rest of the mentioned studies simply classified code comments into SATD comments and non-SATD comments.

There are only two studies that used a different data source than code comments to identify SATD, namely issue tracking systems.
\cite{dai2017detecting} manually examined 8K issues and used the Naive Bayes approach to automatically classify issues into SATD issues and non-SATD issues.
In our previous work \citep{li2022identifying}, we analyzed 23K issue sections (i.e., individual issue summaries, descriptions, or comments) from two issue tracking systems and proposed a convolutional neural network-based approach to identify SATD from those issue sections.

Compared to the aforementioned studies, in this article, we propose an integrated approach to identify four types of SATD (i.e., code/design, requirement, documentation, and test debt) from four different sources (i.e., source code comments, issue trackers, pull requests, and commit messages).
This is the first study that focuses on identifying SATD from multiple sources and is also the first to identify four types of SATD.
Moreover, we present and compare the keywords of different types of SATD and the keywords of SATD from different sources.
Furthermore, we characterize SATD in 103 open-source projects and investigate the relations between SATD in four different sources.

\section{Study Design}
\label{sec:approach}

The goal of this study, formulated according to the Goal-Question-Metric \citep{Solingen:02} template is to ``\textit{\textbf{analyze} data from source code comments, commit messages, pull requests, and issue tracking systems \textbf{for the purpose of} automatically identifying self-admitted technical debt \textbf{with respect to} the identification accuracy, the used keywords in SATD, the quantity of and relations between SATD in different sources \textbf{from the point of view of} software engineers \textbf{in the context of} open-source software.}''
This goal is refined into the following research questions (RQs):%

\begin{itemize}
    \item \textbf{RQ1:}
    \textit{How to accurately identify self-admitted technical debt from different sources?}\\
    \textbf{Rationale:}
    As explained in \cref{sec:introduction}, a fair amount of research has been focused on identifying SATD from source code comments \citep{da2017using,huang2018identifying,ren2019neural,wang2020detecting}.
    However, SATD in issues has hardly been explored \citep{dai2017detecting,li2022identifying}, while SATD identification in pull requests and commit messages has not been investigated before \citep{SIERRA201970}.
    Moreover, there is a lack of integrated approaches to identify SATD from more than one source.
    This research question aims at proposing an approach for SATD identification in different sources with high \emph{accuracy}.
    
    \item \textbf{RQ2:}
    \textit{What are the most informative keywords to identify self-admitted technical debt in different sources?}\\
    \textbf{Rationale:}
    When admitting technical debt in different sources, software engineers potentially have distinct ways of expressing the technical debt.
    For example, developers often write \emph{`TODO'} or \emph{`Fixme'} when admitting technical debt in source code comments, but may not commonly use these terms in other sources.
    Understanding the SATD keywords for different sources could give us an insight into the differences and similarities between sources.
    This can help practitioners identify SATD from different sources using summarized keywords.
    Furthermore, a recent study indicated that the keyword-based SATD identification method achieves a similar or even superior performance for source code comments compared with existing approaches \citep{guo2021far}.
    Thus, extracted keywords could be used to implement light-weighted keyword-based approaches to identify SATD from other sources.
    
    \item \textbf{RQ3:}
    \textit{How much and what types of self-admitted technical debt are documented in different sources?}\\
    \textbf{Rationale:}
    As aforementioned, software engineers can admit technical debt in different sources, but each technical debt item is also of a different type (e.g. design debt, test debt, code debt).
    Quantifying the number of different types of SATD can help to understand how SATD is distributed in different sources and the proportion of types of SATD in different sources.
    The types of SATD could help developers to prioritize SATD; for example, if test debt has higher priority in a company, developers should spend more effort in repaying this type in the sources it is most commonly found. 
    Furthermore, answering this question could help in understanding the advantages and disadvantages of different sources for SATD management.
    For example, if most of the documentation debt is admitted in issue tracking systems, developers can focus on monitoring documentation debt in issues to keep it under control.
    We ask this research question to explore the \emph{characteristics} of SATD in different sources.

    \item \textbf{RQ4:} \textit{What are the relations between self-admitted technical debt in different sources?}\\
    \textbf{Rationale:}
    Previous studies have reported that developers track technical debt using different sources \citep{zampetti2021self}, while different sources are used in different stages during software development \citep{guide1,guide2,guide3}. There are likely interesting relations between SATD in different sources. An example of such a relation was revealed by \cite{zampetti2018self}: SATD which was originally documented in code comments, is sometimes reported as paid back in commit messages.
    Understanding the relations between SATD in different sources can help in understanding the rationale behind admitting technical debt in each of these sources.
    It can also facilitate SATD repayment by grouping related SATD and solving them all together \citep{li2022self}.
    Finally, providing developers with such relations could give them more context to understand the background of the SATD or its possible solutions.
    For example, after discussing the SATD within issues, developers may choose to document it in code comments to be repaid in the future.
    When that time comes, developers can combine the information on the code comments and the discussions in the related issue to make an informed repayment decision.
\end{itemize}

\begin{figure}[thb]
  \centering
  \includegraphics[width=0.5\linewidth]{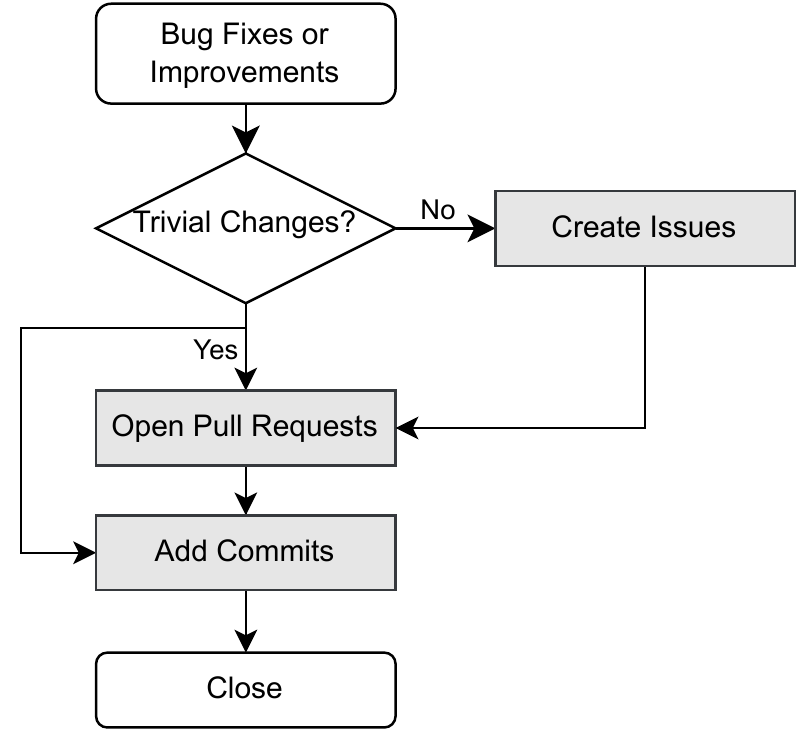}
  \caption{The common workflow.}
  \label{f:flow}
\end{figure}

As mentioned in \cref{sec:introduction}, we choose to identify SATD from code comments, commit messages, pull requests, and issues, as these four sources are the most popular for tracking technical debt among practitioners \citep{zampetti2021self}.
To answer the four stated Research Questions, we need an initial understanding of when and why developers document technical debt in these four sources. To this end, we look into common processes involved in contributing to open-source projects.
According to the contribution guidelines of various Apache projects \citep{guide1,guide2,guide3}, when developers find a bug or want to improve code quality, and that cannot be dealt with trivial changes, they first create an issue to report it, followed by a pull request (see \cref{f:flow}).
If the changes are trivial, some developers choose to create pull requests or even directly push commits to solve them.
Depending on which flow is followed, developers can admit technical debt in any of the involved sources, from just code comments and commits to all four sources. %

\begin{figure}[t]
  \centering
  \includegraphics[width=\linewidth]{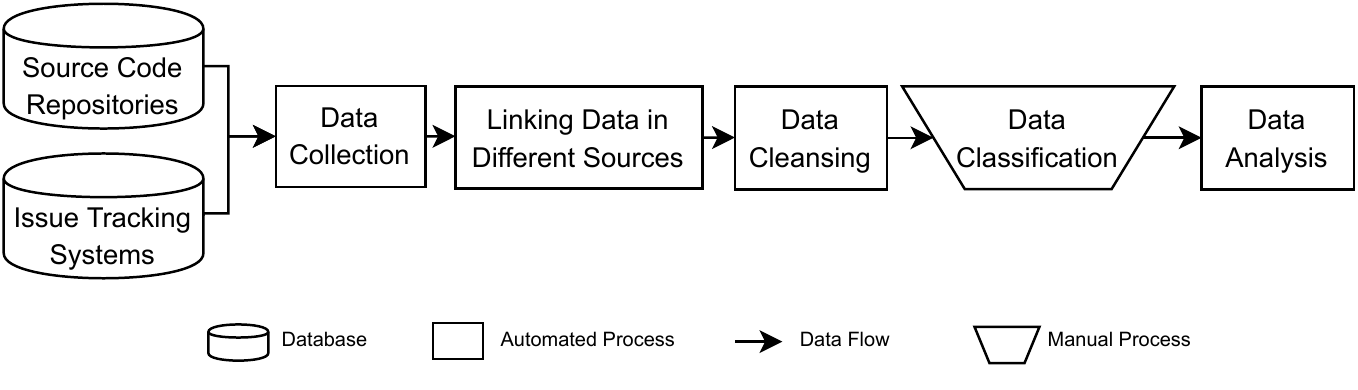}
  \caption{The overview of our approach.}
  \label{f:overview}
\end{figure}

\subsection{Approach Overview}
\label{sec:approach_overview}

The overview of our approach is demonstrated in \cref{f:overview}.
In the first step, we collect pull requests, commit messages, and source code comments from Source Code Repositories and issues from Issue Tracking Systems.
Thereafter, we link the data from different sources for analysis.
Following that, we cleanse, classify, and eventually analyze the data to answer research questions. 
We elaborate on each of the steps in the following subsections.

\begin{table}[b]
\caption{Details of collected data.}
\label{tb:projects}
\begin{center}
\resizebox{\columnwidth}{!}{
\def\arraystretch{1.2}
\begin{tabular}{ccccc|ccccc}
\hline
\multicolumn{5}{c}{\textbf{\# Issues}} & \multicolumn{5}{c}{\textbf{\# Issue Comments}} \\
\hline
Min & Max & Mean & Median & Sum & Min & Max & Mean & Median & Sum \\
\hline
526 & 24,938 & 3,905 & 1,773 & 573,965 & 856 & 303,608 & 21,144 & 5,010 & 3,065,815 \\
\hline
\multicolumn{5}{c}{\textbf{\# Pull Requests}} & \multicolumn{5}{c}{\textbf{\# Pull Comments}} \\
\hline
Min & Max & Mean & Median & Sum & Min & Max & Mean & Median & Sum \\
\hline
507 & 32,072 & 3,035 & 1,608 & 312,591 & 239 & 636,518 & 24,542 & 8,374 & 2,527,803 \\
\hline
\multicolumn{5}{c}{\textbf{\# Commits}} & \multicolumn{5}{c}{\textbf{\# Code Comments}} \\
\hline
Min & Max & Mean & Median & Sum & Min & Max & Mean & Median & Sum \\
\hline
573 & 70,861 & 12,120 & 6,477 & 1,248,324 & 326 & 3,894,056 & 229,705 & 80,211 & 23,659,650 \\
\hline
\end{tabular}
}
\end{center}
\end{table}

\subsection{Data Collection}
\label{sec:data_collection}

To identify SATD in different sources, we first need to find appropriate projects and collect data.
Thus, we look into Apache ecosystems, because these projects are of high quality, maintained by mature communities, and required to make all communications related to code and decision-making publicly accessible \citep{apache2021}.
Since there are over 2,000 repositories in the Apache ecosystem on GitHub\footnote{\url{https://github.com/apache}}, we set the following criteria to select projects pertinent to our study goal:

\begin{enumerate}
    \item The source code repository, commits, pull requests, and issues of the project are publicly available.
    
    \item They have at least 500 issue tickets, 500 pull requests, and 500 commits.
    This ensures that the projects have  sufficient complexity and that we are able to analyze enough projects. We note that, when we try to limit the number of issues and pull requests to 1000, less than 50 projects meet this requirement.
\end{enumerate}

Based on the above criteria, we find 103 Apache projects on GitHub.
The project information was obtained on March 2, 2021.
An overview of the statistics of the four data sources in these projects is presented in \cref{tb:projects}, while the full details are included in the replication package\cref{l:data}.

\begin{figure}[htb]
  \centering
  \includegraphics[width=0.58\linewidth]{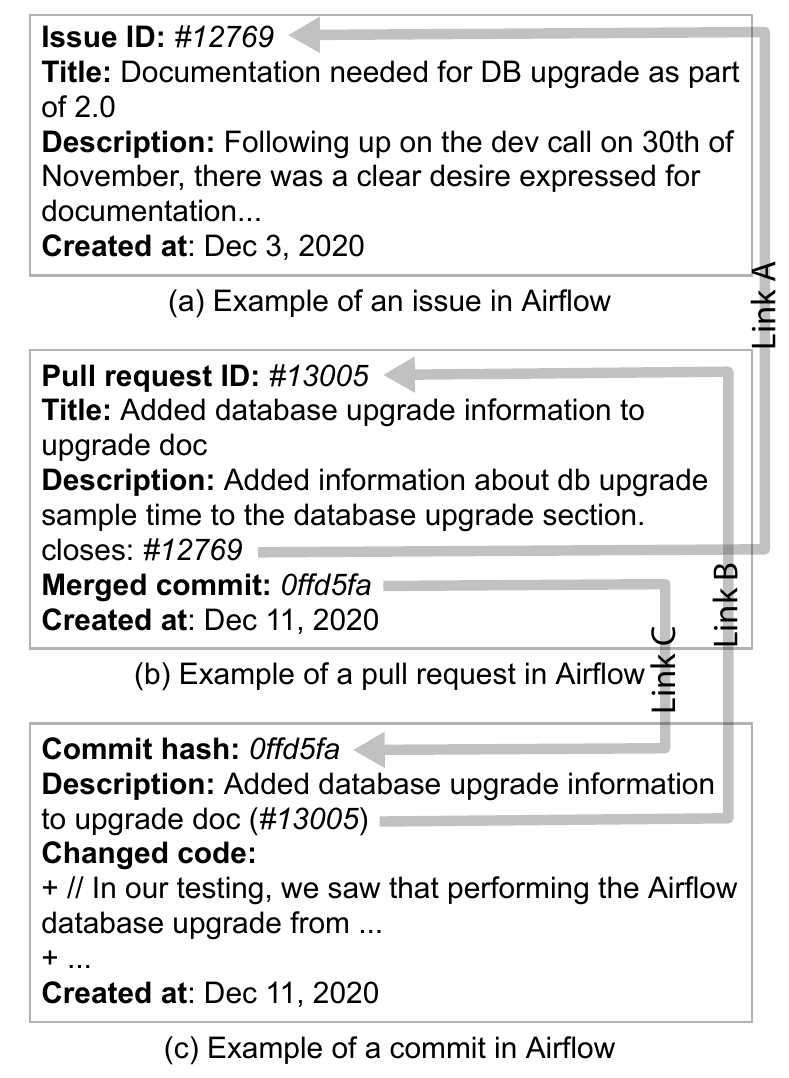}
  \caption{Example of links between different sources in Airflow.}
  \label{f:examples}
\end{figure}

\subsection{Linking Data in Different Sources}
\label{sec:link_sources}

In order to study the relations between SATD in different sources (i.e., answering RQ4), we have to build links between such sources.
Examples of links between different sources are shown in \cref{f:examples} for the Airflow project.
More specifically, since pull request titles or descriptions always contain the issue key information (see \emph{Link A} in \cref{f:examples}), we can build connections between pull requests and issues.
Furthermore, because commit messages contain the related pull request or issue information (see \emph{Link B} in \cref{f:examples}), we can link commits to pull requests or issues.
Moreover, after commits are pushed to the repository, the merged commit hash is also updated in the pull request (see \emph{Link C} in \cref{f:examples}), thus closing the loop between pull requests and commits.
Finally, commits record changes to one or more files, thus we can link code comment changes with commits.

\subsection{Data Cleansing}

In issues and pull requests, apart from comments left by developers, plenty of comments are automatically generated by bots.
For example, when a new contributor opens a first pull request, a welcome bot could comment on a brief contribution guideline to help the new contributor.
Since comments created by bots do not contain SATD, we filtered these comments out.
Specifically, we gathered the 100 most active users by the number of submitted comments.
Then we checked comments posted by these users, identified the bot users (e.g., \textit{Hadoop QA} and \textit{Hudson}), and removed all comments posted by bot users according to the list of bots' usernames.

Moreover, developers sometimes copy and paste code snippets in different sources for various reasons. Because we focus on the technical debt admitted by developers instead of the source code, we removed source code in code blocks.

Finally, before feeding the data into machine learning models, we also removed numbers, removed non-English characters, and converted the remaining data into lowercase.

\begin{table}[t]
\caption{Types and indicators of self-admitted technical debt.}
\label{tb:statistics_type_projects}
\begin{center}
\resizebox{\columnwidth}{!}{
\def\arraystretch{1.2}
\begin{tabular}{@{\extracolsep{4pt}}L{0.9cm}L{2.2cm}L{8.8cm}@{}@{}}
\hline
\textbf{Type} & \textbf{Indicator} & \textbf{Definition} \\
\hline
Arch. & Violation of modularity & Because shortcuts were taken, multiple modules became inter-dependent, while they should be independent. \\
 & Using obsolete technology & Architecturally-significant technology has become obsolete. \\
\hline
Build & Over- or under-declared dependencies & Under-declared dependencies: dependencies in upstream libraries are not declared and rely on dependencies in lower-level libraries. Over-declared dependencies: unneeded dependencies are declared. \\
 & Poor deployment practice & The quality of deployment is low that compile flags or build targets are not well organized. \\
\hline
Code & Complex code & Code has accidental complexity and requires extra refactoring action to reduce this complexity. \\
 & Dead code & Code is no longer used and needs to be removed. \\
 & Duplicated code & Code that occurs more than once instead of as a single reusable function. \\
 & Low-quality code & Code quality is low, for example, because it is unreadable, inconsistent, or violating coding conventions. \\
 & Multi-thread correctness & Thread-safe code is not correct and may potentially result in synchronization problems or efficiency problems. \\
 & Slow algorithm & A non-optimal algorithm is utilized that runs slowly. \\
\hline
Defect & Uncorrected known defects & Defects are found by developers but ignored or deferred to be fixed. \\
\hline
Design & Non-optimal decisions & Non-optimal design decisions are adopted. \\
\hline
Doc. & Low-quality documentation & The documentation has been updated reflecting the changes in the system, but the quality of updated documentation is low. \\
 & Outdated documentation & A function or class is added, removed, or modified in the system, but the documentation has not been updated to reflect the change. \\
\hline
Req. & Requirements partially implemented & Requirements are implemented, but some are not fully implemented. \\
 & Non-functional requirements not being fully satisfied & Non-functional requirements (e.g. availability, capacity, concurrency, extensibility), as described by scenarios, are not fully satisfied. \\
\hline
Test & Expensive tests & Tests are expensive, resulting in slowing down testing activities. Extra refactoring actions are needed to simplify tests. \\
 & Flaky tests & Tests fail or pass intermittently for the same configuration. \\
 & Lack of tests & A function is added, but no tests are added to cover the new function. \\
 & Low coverage & Only part of the source code is executed during testing. \\
\hline
\end{tabular}
}
\end{center}
\end{table}

\subsection{Data Classification}
\label{sec:data_classification}

As mentioned in \cref{sec:introduction}, there are no SATD datasets available for commit messages and pull requests \citep{SIERRA201970}.
We thus need to manually analyze commits and pull requests to create the datasets for training machine learning models.
In our previous work \citep{li2020identification}, we found out that developers commonly share a single type of SATD in an issue section: either an issue description or a comment, as a whole.
Moreover, other researchers \citep{maldonado2015detecting,da2017using} assigned a single type of SATD to a whole source code comment. 
Therefore, we decided to classify whole sections, rather than individual sentences.
Specifically, a pull request is typically composed of a pull request summary, a pull request description, and a number of normal and code review comments.
Thus, similarly to our previous study \citep{li2022identifying}, we call each part of a pull request (i.e., summary, description, or comment) a \emph{pull request section}, and analyzed them separately.
Because commit messages cannot be divided, we classify a whole commit message as different types of SATD or non-SATD.

Since our previous work reported that 3,400 pieces of data are sufficient for a similar SATD identification task \citep{li2022identifying} and the cost of manual analysis is high, we decided to analyze 5,000 items for both commit messages and pull request sections.

We treated each commit message and pull request section individually and classified them as different types of SATD or non-SATD according to the classification framework proposed by \cite{li2022identifying}.
The definitions of the different types of SATD from \cite{li2022identifying} are shown in \cref{tb:statistics_type_projects}.
Specifically, we used an open-source text annotation tool (Doccano\footnote{\url{https://github.com/doccano/doccano}}) to annotate different types of SATD or non-SATD for each commit message and pull request section.
To ensure that the first and second authors can reach an agreement on classification, they manually classified 50 commits or pull sections (according to the classification framework in \cref{tb:statistics_type_projects}) independently.
These results were compared and discussed between the two authors to make sure that the annotations of each item align with the definitions of the types of SATD in the classification framework.
This process was repeated two times to ensure a better and more uniform understanding of the types of SATD.
After all the commit messages and pull request sections were classified by the first author, we randomly selected a sample of this data with a size greater than the statistically significant sample size (i.e., 372).
Then the second author independently classified the sample, and Cohen's kappa coefficient \citep{landis1977measurement} was calculated. This coefficient measures the level of agreement between the classifications of the first author and the second author. It is commonly used to measure inter-rater reliability, and can thus be useful when determining the risk of bias in the reliability of the classification.
The results indicate that we have achieved `substantial' agreement \citep{landis1977measurement} with the coefficient of $+0.74$.

\begin{table}[htb]
\caption{Number of different types of SATD.}
\label{tb:satd_num_analyzed}
\begin{center}
\resizebox{\columnwidth}{!}{
\def\arraystretch{1.2}
\begin{tabular}{ccccc}
\hline & \\[-2.5ex]
\multirow{2}{*}{\textbf{\makecell[c]{Type}}} & \multicolumn{4}{c}{\textbf{Source}} \\
\cline{2-5}
 & Code Comment & Issue Section & Pull Section & Commit Message \\
\hline
Code/Design Debt & 2,703 & 2,169 & 510 & 522 \\
Documentation Debt & 54 & 487 & 101 & 98 \\
Test Debt & 85 & 338 & 68 & 58 \\
Requirement Debt & 757 & 97 & 20 & 27 \\
Other & 58,676 & 20,089 & 4,301 & 4,295 \\
\hline
\end{tabular}
}
\end{center}
\end{table}

It is noted that the SATD dataset in code comments does not differentiate between code debt and design debt because the similarity between them is high \citep{da2017using}.
Thus, in this work, we combined these two types of SATD when training and evaluating SATD classifiers.
Besides, we chose to identify four types of SATD, namely \emph{code/design debt}, \emph{documentation debt}, \emph{test debt}, and \emph{requirement debt}, because these four types of SATD are prevalent in the four data sources we analyzed \citep{da2017using,li2022identifying}.
The number of different types of SATD items for training and testing machine learning models is presented in \cref{tb:satd_num_analyzed}.

\subsection{Data Analysis}%

\subsubsection{Machine Learning Models:}
\label{sec:ml}

Because there is no approach designed to identify SATD from different sources, inspired by the work of \cite{kim2014convolutional} and \cite{liu2015representation}, we propose Multitask Text Convolutional Neural Network (MT-Text-CNN) to fill this gap.
More specifically, because Text-CNN has been proven to be efficient in SATD identification in previous work \citep{ren2019neural,li2022identifying}, we thus leverage the multitask learning technique \citep{liu2015representation} in combination with Text-CNN and then propose our approach.
As mentioned in \cref{sec:introduction}, there are three advantages to using the multitask learning technique to detect SATD from different sources: 1) MT-Text-CNN can be used to identify SATD from multiple sources, compared to training and using multiple machine learning models; 2) MT-Text-CNN is more extensible and can thus support identifying SATD from new sources more easily; 3) MT-Text-CNN can learn to identify SATD from different sources in parallel while learning a shared representation, which could potentially improve the predictive performance of identifying SATD from individual sources.
In order to evaluate the predictive performance of our approach when identifying SATD from different sources, we compare its performance with several machine learning approaches in SATD identification.
All used machine learning approaches are listed below:

\begin{itemize}
    \item \textbf{Traditional machine learning approaches (LR, SVM, RF)}:
    To illustrate the effectiveness of our approach, we select and compare our approach with three prevalent traditional machine learning algorithms, namely Logistic Regression (LR) \citep{genkin2007large}, Support Vector Machine (SVM) \citep{sun2009strategies}, and Random Forest (RF) \citep{breiman2001random}.
    We use \textit{TF-IDF} to vectorize the input data and train these three traditional classifiers using the implementation in Sklearn\footnote{\label{l:sklearn}\url{https://scikit-learn.org}} with default settings.

    \begin{figure}[t]
      \centering
      \includegraphics[width=\linewidth]{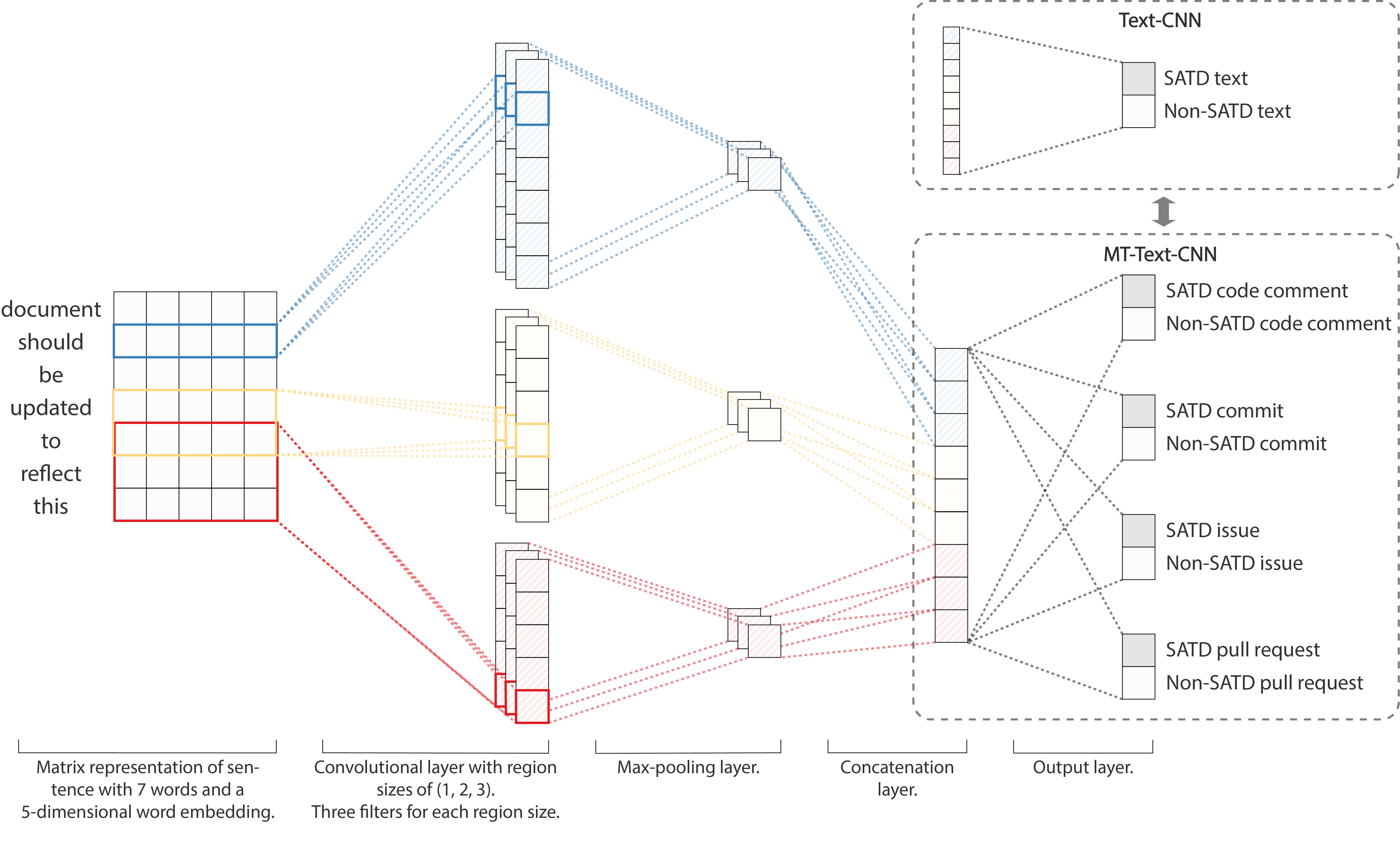}
      \caption{Architectures of Text-CNN and MT-Text-CNN for classifying SATD and non-SATD.}
      \label{f:cnn_structure}
    \end{figure}
    
    \item \textbf{Text Convolutional Neural Network (Text-CNN)}:
    Text-CNN is a state-of-the-art text classification algorithm proposed by \cite{kim2014convolutional}, which has been used in several SATD identification studies \citep{ren2019neural,li2022identifying}.
    The details of this approach are given, as they are background knowledge for understanding the differences between Text-CNN and MT-Text-CNN.
    The architecture of Text-CNN is demonstrated in \cref{f:cnn_structure}.
    As can be seen, Text-CNN consists of five layers, namely the embedding layer, convolutional layer, max-pooling layer, concatenation layer, and output layer.
    
    \begin{itemize}
        \item \textbf{Embedding layer}:
        It is the first layer that converts the tokenized input sentence (the length of the sentence is $n$) into a matrix of size $n \times k$ using a k-dimensional word embedding (see \cref{sec:word_embed}).
        For example in \cref{f:cnn_structure}, the input sentence is \textit{document should be updated to reflect this}, which is transformed into a $7 \times 5$ matrix as the input sentence contains 7 words and the word embedding dimensionality equals to 5.
        
        \item \textbf{Convolutional layer}:
        It is the fundamental layer of CNN that performs convolution operations to extract the high-level features from the sentence matrix.
        A convolution operation associates a filter, which is a matrix that has the same width as the sentence matrix (i.e., $k$) and the height of it varies.
        The height of the filter is denoted by region size.
        The filter with a region size of $h$ can be applied to a window of $h$ words to generate a new feature.
        Thus, by sliding a filter with a region size of $h$ over the whole sentence matrix, a feature map of size $n - h + 1$ is produced.
        For instance in \cref{f:cnn_structure}, when the model has filters whose region sizes are 1, 2, and 3, the sizes of produced feature maps are 7, 6, and 5 respectively. 
        
        \item \textbf{Max-pooling layer}:
        It is a layer that calculates the maximum value of each feature map to reduce the spatial size of the representation.
        
        \item \textbf{Concatenation layer}:
        It is a layer that concatenates the scalar features to form the penultimate layer.
        
        \item \textbf{Output layer}:
        It is the last layer that computes the probability of input text to be SATD text.
        Because Text-CNN is for a single task, it performs a linear transformation of the features from the previous layer by $ Y = \textbf{W} \cdot X + B $, where $\textbf{W}$ and $B$ denotes weight and bias.
        The length of $Y$ equals the number of classes.
        Then the \emph{softmax function} is applied to $Y$ to calculate the probability of input text belonging to each class.
        For example, there are two classes: SATD text and non-SATD text in \cref{f:cnn_structure}.
        In this work, because we focus on identifying different types of SATD or non-SATD, we have five classes: four types of SATD text and non-SATD text.
    \end{itemize}
    
    \item \textbf{Multitask Text Convolutional Neural Network (MT-Text-CNN)}:
    Although SATD in different sources has substantial similarities, there still are significant differences between them \citep{li2022identifying}.
    This could lower the accuracy of Text-CNN when detecting SATD from multiple sources, as the standards for SATD identification are slightly different for different sources.
    Thus, we propose the MT-Text-CNN approach to accurately identify SATD from different sources.
    The architecture of MT-Text-CNN is illustrated in \cref{f:cnn_structure}.
    As we can see, apart from the output layer, the rest of the layers are identical to Text-CNN.
    Inspired by the work of \cite{liu2015representation}, for each task, we create a task-specific output layer, which also performs a linear transformation of the features from the previous layer by $ Y^{(t)} = \textbf{W}^{(t)} \cdot X + B^{(t)} $, where $t$ denotes different tasks (i.e., identifying SATD from different sources).
    Then the \emph{softmax function} is applied to $Y^{(t)}$ to calculate the probability of input text belonging to each class for task $t$.
\end{itemize}

In this study, we implement machine learning approaches using Pytorch library\footnote{\url{https://pytorch.org}}.
Machine learning models are trained on NVIDIA Tesla V100 GPUs.

\subsubsection{Baseline Approaches:}

We implement one baseline approach to compare the results with machine learning approaches.

\begin{itemize}
    \item \textbf{Random Classifier (Random)}:
    It classifies text as SATD randomly according to the probability of random text being SATD text.
    For instance, if the database contains 1,000 pieces of SATD text out of 10,000 pieces of text, this approach assumes the probability of new text to be SATD text is $1000 / 10000 = 10\%$.
    Then this approach randomly classifies any text as SATD text corresponding to the calculated probability (10\%).
\end{itemize}

\subsubsection{Word Embedding:}
\label{sec:word_embed}

Word embedding is a type of word representation where words are represented similarly if they have high similarities.
Typically, words are represented in the form of real number vectors.
Training word embedding using data in the same context of the target task has been proven to outperform randomly initialized or pre-trained word embeddings by a large margin for SATD identification task \citep{li2022identifying}.
In this study, we train word embedding on our collected data (i.e., source code comments, commit messages, pull requests, and issues) using the fastText technique \citep{mikolov2018advances} while setting the word embedding dimension to 300.

\subsubsection{Model Evaluation Approach:}

To reduce the bias during training and testing, we used stratified 10-fold cross-validation to evaluate the predictive performance of SATD identification approaches.
Specifically, this approach randomly splits the dataset into 10 partitions while ensuring all partitions have the same proportion of different types of SATD.
After that, for each partition, we use it as test data and take the remaining nine partitions as the training dataset to train the model.
Finally, we calculate the average predictive performance of the model using the obtained evaluation scores.

\subsubsection{Multitask Network Training Procedure:}

We follow the guideline proposed by Collobert and Weston \citep{collobert2008unified} to perform joint training on multiple tasks.
Training is done in a stochastic manner with the following steps:

\begin{itemize}
    \item Randomly pick up a task.
    \item Get a random training sample for this task.
    \item Train the machine learning model using the sample.
    \item Go to the first step.
\end{itemize}

\subsubsection{Strategies for Handling Imbalanced Data:}
\label{sec:imbalanced_data}

According to the previous study \citep{li2022identifying,ren2019neural}, only a very small proportion of source code comments or issue sections are SATD comments, so the dataset is seriously imbalanced.
It has been shown that using weighted loss could effectively improve the SATD identification accuracy \citep{li2022identifying,ren2019neural}, which penalizes harder the wrongly classified items from minority classes (i.e., false negative and false positive errors) during training.
Thus, we use the weighted loss function in this work.

\subsubsection{Evaluation Metrics:}

We use the following statistics: true positive (TP) represents the number of items correctly classified as SATD items; true negative (TN) represents the number of items correctly classified as non-SATD items; false positive (FP) represents the number of items that are wrongly classified as SATD items; false negative (FN) represents the number of items that are wrongly classified as non-SATD items.
Sequentially, we calculate \textbf{precision} ($\frac{TP}{TP+FP}$), \textbf{recall} ($\frac{TP}{TP+FN}$), and \textbf{F1-score} ($2 \times \frac{precision \times recall}{precision + recall}$) to evaluate the performance of different approaches.
High evaluation metrics indicate good performance.
We use F1-score to evaluate the performance of approaches because it incorporates the trade-off between precision and recall.
It is noted that when identifying different types of SATD, we first calculate the F1-score for each type of SATD, and then average the F1-score to obtain the macro F1-score.

\subsubsection{Keyword Extraction:}
\label{sec:keyword_extraction}

To extract keywords that indicate SATD (to answer RQ2), we utilize the approach introduced by \cite{ren2019neural}.
This method extracts keywords by finding the most important features based on the trained Text-CNN model using the backtracking technique.
Specifically, as shown in \cref{f:cnn_structure}, this approach multiples the results of the concatenation layer by the weights of the output layer to find features that contribute most to the classification.
Then it locates the text phrases that are related to the important features using backtracking.
After that, we can summarize SATD keywords based on the extracted text phrases.

\subsubsection{SATD Similarity Calculation:}
\label{sec:satd_similarity}

To understand the relations between SATD in different sources (to answer RQ4), we calculate the cosine similarity between SATD items from different sources.
We choose cosine similarity similarly to a previous study linking SATD in comments and commits \citep{zampetti2018self}.
Specifically, we preprocess SATD items by removing the numbers, converting them to lowercase, and removing stop words.
After that, we deploy Bag-Of-Words (BoW) to generate vectors from input text.
Then we calculate the cosine similarity using the \emph{Scipy}\footnote{\url{https://scipy.org}} package.

\section{Results}
\label{sec:results}

\begin{table}[t]
\caption{Comparison of F1-score between machine learning and baseline approaches.}
\label{tb:f1_all_methods}
\begin{center}
\resizebox{\columnwidth}{!}{
\def\arraystretch{1.2}
\begin{tabular}{ccccccccc}
\hline & \\[-2.5ex]
\multicolumn{2}{c}{\multirow{4}{*}{\textbf{Classifier}}} & \multirow{4}{*}{\textbf{\makecell{Type of\\SATD}}} & \multicolumn{5}{c}{\multirow{3}{*}{\textbf{Source}}} & \multirow{4}{*}{\textbf{\makecell{Average\\Imp.\\over\\Random}}} \\
 & & & & & & & & \\
 & & & & & & & & \\
\cline{4-8}
 & & & Comment & Commit & Pull & Issue & Avg. & \\
\hline
\parbox[t]{2mm}{\multirow{10}{*}{\rotatebox[origin=c]{90}{\textbf{Deep Learning}}}} & \multirow{5}{*}{\textbf{Text-CNN}} & C/D. & 0.665 & 0.485 & 0.515 & 0.461 & 0.531 & 7.8$\times$ \\
 & & DOC. & 0.526 & 0.632 & 0.484 & 0.456 & 0.524 & 52.4$\times$ \\
 & & TST. & 0.443 & 0.469 & 0.507 & 0.463 & \textbf{0.471} & 157.0$\times$ \\
 & & REQ. & 0.566 & 0.217 & 0.299 & 0.343 & 0.356 & 71.2$\times$ \\
\cline{3-8}
\cline{9-9}
 & & AVG. & 0.550 & 0.451 & \textbf{0.451} & 0.431 & 0.471 & 22.4$\times$ \\
\cline{2-9}
 & \multirow{5}{*}{\textbf{MT-Text-CNN}} & C/D. & 0.725 & 0.536 & 0.539 & 0.486 & \textbf{0.571} & 8.4$\times$ \\
 & & DOC. & 0.626 & 0.659 & 0.441 & 0.457 & \textbf{0.546} & 54.6$\times$ \\
 & & TST. & 0.540 & 0.449 & 0.461 & 0.432 & 0.470 & 159.7$\times$ \\
 & & REQ. & 0.585 & 0.255 & 0.325 & 0.437 & \textbf{0.400} & 80.0$\times$ \\
\cline{3-8}
\cline{9-9}
 & & AVG. & \textbf{0.619} & \textbf{0.475} & 0.441 & \textbf{0.453} & \textbf{0.497} & 23.7$\times$ \\
\hline
\parbox[t]{2mm}{\multirow{15}{*}{\rotatebox[origin=c]{90}{\textbf{Traditional Machine Learning}}}} & \multirow{5}{*}{\textbf{\makecell{LR}}} & C/D. & 0.613 & 0.327 & 0.457 & 0.353 & 0.438 & 6.4$\times$ \\
 & & DOC. & 0.352 & 0.556 & 0.281 & 0.235 & 0.356 & 35.6$\times$ \\
 & & TST. & 0.245 & 0.129 & 0.206 & 0.228 & 0.202 & 67.3$\times$ \\
 & & REQ. & 0.389 & 0.000 & 0.000 & 0.019 & 0.102 & 20.4$\times$ \\
\cline{3-8}
\cline{9-9}
 & & AVG. & 0.400 & 0.253 & 0.236 & 0.208 & 0.274 & 13.0$\times$ \\
\cline{2-9}
 & \multirow{5}{*}{\textbf{SVM}} & C/D. & 0.400 & 0.051 & 0.085 & 0.008 & 0.136 & 2.0$\times$ \\
 & & DOC. & 0.085 & 0.566 & 0.202 & 0.094 & 0.237 & 23.7$\times$ \\
 & & TST. & 0.000 & 0.074 & 0.038 & 0.067 & 0.045 & 15.0$\times$ \\
 & & REQ. & 0.200 & 0.000 & 0.000 & 0.000 & 0.050 & 10.0$\times$ \\
\cline{3-8}
\cline{9-9}
 & & AVG. & 0.171 & 0.173 & 0.081 & 0.042 & 0.117 & 5.6$\times$ \\
\cline{2-9}
 & \multirow{5}{*}{\textbf{RF}} & C/D. & 0.600 & 0.199 & 0.095 & 0.065 & 0.240 & 3.5$\times$ \\
 & & DOC. & 0.500 & 0.630 & 0.240 & 0.092 & 0.366 & 36.6$\times$ \\
 & & TST. & 0.289 & 0.124 & 0.101 & 0.119 & 0.158 & 52.7$\times$ \\
 & & REQ. & 0.584 & 0.000 & 0.000 & 0.056 & 0.160 & 32.0$\times$ \\
\cline{3-8}
\cline{9-9}
 & & AVG. & 0.494 & 0.238 & 0.109 & 0.083 & 0.231 & 11.0$\times$ \\
\hline
\parbox[t]{2mm}{\multirow{5}{*}{\rotatebox[origin=c]{90}{\textbf{Baseline}}}} & \multirow{5}{*}{\textbf{Random}} & C/D. & 0.053 & 0.071 & 0.071 & 0.076 & \underline{0.068} & \\
 & & DOC. & 0.001 & 0.003 & 0.021 & 0.013 & \underline{0.010} & \\
 & & TST. & 0.002 & 0.004 & 0.000 & 0.005 & \underline{0.003} & \\
 & & REQ. & 0.009 & 0.000 & 0.004 & 0.005 & \underline{0.005} & \\
\cline{3-8}
 & & AVG. & \underline{0.016} & \underline{0.020} & \underline{0.024} & \underline{0.025} & \underline{0.021} & \\
\hline
\end{tabular}
}
\end{center}
\end{table}

\subsection{(RQ1) \textit{How to accurately identify self-admitted technical debt from different sources?}}
\label{sec:rq1}

In order to accurately identify SATD from multiple sources, we have proposed a deep learning approach, namely MT-Text-CNN which leverages the concepts of CNN networks and multi-task learning (see details in \cref{sec:ml}).

\subsubsection{Comparing the predictive performance of different classifiers.}

To evaluate the effectiveness of our approach, we first compare our approach with a prevalent deep learning approach (i.e., Text-CNN), three traditional machine learning approaches (i.e., LR, SVM, and RF), and one baseline method (i.e., random classifier).
We train MT-Text-CNN and Text-CNN with randomized word vectors using the same default hyperparameter settings of Text-CNN \citep{kim2014convolutional}.
Because a class imbalance problem was identified in two previous studies that concerned a SATD source code comment dataset \citep{ren2019neural} and a SATD issue dataset \citep{li2022identifying}, we use stratified 10-fold cross-validation to eliminate potential bias caused by this problem when evaluating the aforementioned approaches.
\cref{tb:f1_all_methods} presents the F1-score of the aforementioned approaches identifying different types of SATD (i.e., code/design debt, documentation debt, test debt, and requirement debt) from different sources (i.e., code comment, commit message, pull request, and issue tracker) as well as the F1-score comparison between machine learning approaches and the baseline.
It is noted that C/D., DOC., TST., and REQ. refers to code/design debt, documentation debt, test debt, and requirement debt, respectively.
Furthermore, the best F1-score is highlighted in bold while the worst is underlined.
As we can see in \cref{tb:f1_all_methods}, the two deep learning approaches (i.e., MT-Text-CNN and Text-CNN) achieve a significantly higher average F1-score compared to other approaches.
Moreover, our MT-Text-CNN approach achieves the highest average F1-score of 0.497, outperforming Text-CNN with respect to both the average F1-score across sources (ranging between 0.441 to 0.619 versus 0.431 to 0.550) and the average F1-score across different types of SATD (ranging between 0.400 to 0.571 versus 0.356 to 0.531).
Furthermore, we perform T-Test \citep{student1908probable} and Cliff's delta \citep{vargha2000critique} to evaluate the differences between the F1-score of MT-Text-CNN and other approaches; the results are presented in \cref{tb:p_value}.
It is noted that effect sizes are marked as \emph{small} ($0.11 \le d < 0.28$), \emph{medium} ($0.28 \le d < 0.43$), and \emph{large} ($0.43 \le d$) based on suggested benchmarks \citep{vargha2000critique}.
We can observe that the effect sizes between MT-Text-CNN and others are all categorized as \emph{large}.
The results indicate that the differences between the F1-score of MT-Text-CNN and other approaches are statistically significant as all the \textit{p}-values are less than $0.05$.
In comparison, the average F1-score obtained by traditional machine learning approaches ranges from 0.117 to 0.231, while the random method achieves the lowest F1-score of 0.021.

\begin{table}[t]
\caption{Comparison of the F1-score between MT-Text-CNN and other approaches.}
\label{tb:p_value}
\begin{center}
\resizebox{0.85\columnwidth}{!}{
\def\arraystretch{1.2}
\begin{tabular}{cccccc}
\hline & \\[-2.5ex]
\textbf{Approach} & \textbf{Text-CNN} & \textbf{LR} & \textbf{SVM} & \textbf{RF} & \textbf{Random} \\
\hline
\textit{p}-value & 3.62e-05 & 5.54e-59 & 1.37e-53 & 2.19e-32 & 2.49e-104 \\
Cliff's delta & 0.44 (large) & 1.0 (large) & 1.0 (large) & 1.0 (large) & 1.0 (large) \\
\hline
\end{tabular}
}
\end{center}
\end{table}

\begin{framed}
\noindent \textit{Our MT-Text-CNN approach achieved the highest average F1-score of 0.497 when identifying four types of SATD from multiple sources (i.e., comments, commits, pull requests, and issues).}
\end{framed}

\subsubsection{Improving the MT-Text-CNN approach.}

To further improve the predictive performance of our proposed approach, we investigate word embedding configurations, strategies to handle imbalanced data, and hyper-parameter tuning (see details in \cref{sec:word_embed,sec:imbalanced_data}).

First, we improve the word embeddings by training it on our collected data (i.e., source code comments, commit messages, pull requests, and issues) using the fastText technique, and compare the results with the randomized word embeddings.
As can be seen in \cref{tb:word_embedding}, using the trained word embeddings significantly improved the F1-score compared to the randomly initialized word embedding.
It is noted that enabling fine-tuning word embedding during training (i.e., setting the word embedding to non-static) achieved a worse F1-score compared to using the static trained word embedding (0.524 versus 0.549).
Therefore, we chose to use trained word embedding while setting it to static during training.

\begin{table}[thb]
\caption{Comparison of F1-score between different word embedding configurations.}
\label{tb:word_embedding}
\begin{center}
\resizebox{0.95\columnwidth}{!}{
\def\arraystretch{1.2}
\begin{tabular}{ccccccc}
\hline & \\[-2.5ex]
\multirow{2}{*}{\textbf{Word Embedding}} & \multirow{2}{*}{\textbf{Dimensions}} & \multicolumn{5}{c}{\textbf{Source}} \\
\cline{3-7}
 & & Comment & Commit & Pull & Issue & Avg. \\
\hline
Random (non-static) & 300 & 0.619 & 0.475 & 0.441 & 0.453 & \underline{0.497} \\
Trained (non-static) & 300 & 0.612 & 0.521 & 0.496 & 0.469 & 0.524 \\
Trained (static) & 300 & 0.652 & 0.564 & 0.470 & 0.509 & \textbf{0.549} \\
\hline
\end{tabular}
}
\end{center}
\end{table}

Second, SATD datasets commonly have the issue of imbalanced data (i.e., the percentage of SATD comments is significantly less than non-SATD comments). Thus, we improve the predictive performance using weighted loss to eliminate the influence of imbalanced data, which has been shown to be an efficient approach to deal with imbalanced data in the previous work (see \cref{sec:imbalanced_data}).
In \cref{tb:weigthted_loss}, we can observe that the F1-score is improved from 0.549 to 0.593 by applying the weighted loss compared to the default loss function. 
Thus, we adopt weighted loss to mitigate the effects of imbalanced datasets.

\begin{table}[thb]
\caption{Comparison of F1-score between default loss function and weighted loss function.}
\label{tb:weigthted_loss}
\begin{center}
\resizebox{.75\columnwidth}{!}{
\def\arraystretch{1.2}
\begin{tabular}{cccccc}
\hline & \\[-2.5ex]
\multirow{2}{*}{\textbf{Type}} & \multicolumn{5}{c}{\textbf{Source}} \\
\cline{2-6}
 & Comment & Commit & Pull & Issue & Avg. \\
\hline
Default & 0.652 & 0.564 & 0.470 & 0.509 & \underline{0.549} \\
Weighted loss & 0.642 & 0.612 & 0.573 & 0.544 & \textbf{0.593} \\
\hline
\end{tabular}
}
\end{center}
\end{table}

\begin{table}[b]
\caption{Comparison of F1-score between different combinations of region sizes.}
\label{tb:parameter_tunning}
\begin{center}
\resizebox{0.85\columnwidth}{!}{
\def\arraystretch{1.2}
\begin{tabular}{ccccccc}
\hline & \\[-2.5ex]
\multicolumn{2}{c}{\multirow{2}{*}{\textbf{Region Size}}} & \multicolumn{5}{c}{\textbf{Source}} \\
\cline{3-7}
 & & Comment & Commit & Pull & Issue & Avg. \\
\hline
\parbox[t]{2mm}{\multirow{4}{*}{\rotatebox[origin=c]{90}{\textbf{Single}}}} & (1) & 0.553 & 0.551 & 0.513 & 0.454 & 0.518 \\
 & (3) & 0.602 & 0.643 & 0.541 & 0.523 & \textbf{0.577} \\
 & (5) & 0.593 & 0.582 & 0.518 & 0.491 & 0.546 \\
 & (7) & 0.567 & 0.513 & 0.477 & 0.451 & \underline{0.502} \\
\hline
\parbox[t]{2mm}{\multirow{8}{*}{\rotatebox[origin=c]{90}{\textbf{Multiple}}}} & (1,2,3) & 0.632 & 0.647 & 0.596 & 0.596 & 0.606 \\
 & (2,3,4) & 0.644 & 0.645 & 0.573 & 0.553 & 0.604 \\
 & (3,4,5) & 0.642 & 0.612 & 0.573 & 0.544 & 0.593 \\
 & (1,2,3,4) & 0.662 & 0.640 & 0.574 & 0.557 & 0.608 \\
 & (1,3,5,7) & 0.652 & 0.631 & 0.559 & 0.541 & 0.596 \\
 & (2,4,6,8) & 0.644 & 0.612 & 0.571 & 0.542 & \underline{0.592} \\
 & (1,2,3,4,5) & 0.656 & 0.642 & 0.581 & 0.555 & \textbf{0.609} \\
 & (1,2,3,4,5,6) & 0.664 & 0.626 & 0.574 & 0.559 & 0.606 \\
 & (1,2,3,4,5,6,7) & 0.662 & 0.615 & 0.576 & 0.558 & 0.603 \\
\hline
\end{tabular}
}
\end{center}
\end{table}

Third, we follow the guideline proposed by \cite{zhang2017sensitivity} to fine-tune the hyper-parameters of our neural network.
Specifically, we conducted a line search over the single filter region size (i.e., using (1), (3), (5), and (7) as the region size) for the best single region size.
As shown in \cref{tb:parameter_tunning}, the single filter size (3) is the best for SATD identification.
After that, we investigated the effectiveness of combining multiple filters whose region sizes are close to the best single region size (3).
Because we cannot explore all the combinations of region sizes, we tested the following multiple region sizes: (1,2,3), (2,3,4), (3,4,5), (1,2,3,4), (1,3,5,7), (2,4,6,8), (1,2,3,4,5), (1,2,3,4,5,6), and (1,2,3,4,5,6,7).
The F1-score of each multiple filter's configurations is shown in \cref{tb:parameter_tunning}.
As we can see, all combinations of multiple filters outperform the F1-score of the best single region size (3), while the region size of (1,2,3,4,5) achieved the best F1-score of 0.609.
Thus, we use (1,2,3,4,5) as the region sizes for our approach.

Lastly, we explore the effect of the number of feature maps for each filter region size.
According to the guideline \citep{zhang2017sensitivity}, we explored the number of feature maps from 50 to 800.
Observing \cref{tb:feature_number}, using 200 feature maps achieves the best average F1-score of 0.611.

\begin{table}[thb]
\caption{Comparison of F1-score between the different number of feature maps.}
\label{tb:feature_number}
\begin{center}
\resizebox{0.75\columnwidth}{!}{
\def\arraystretch{1.2}
\begin{tabular}{cccccc}
\hline & \\[-2.5ex]
\multirow{2}{*}{\textbf{\makecell[c]{Number of\\Features}}} & \multicolumn{5}{c}{\textbf{Source}} \\
\cline{2-6}
 & Comment & Commit & Pull & Issue & Avg. \\
\hline
50 & 0.645 & 0.643 & 0.563 & 0.551 & 0.601 \\
100 & 0.656 & 0.642 & 0.581 & 0.555 & 0.609 \\
200 & 0.666 & 0.644 & 0.578 & 0.557 & \textbf{0.611} \\
400 & 0.650 & 0.638 & 0.558 & 0.558 & 0.601 \\
800 & 0.634 & 0.639 & 0.546 & 0.550 & \underline{0.592} \\
\hline
\end{tabular}
}
\end{center}
\end{table}

\begin{framed}
\noindent \textit{The average F1-score of our MT-Text-CNN approach to detect four types of SATD from multiple sources is improved from 0.497 to 0.611 by 22.9\% after word embedding improvement, imbalanced data handling, and hyper-parameter tuning.}
\end{framed}

\subsection{(RQ2) \textit{What are the most informative keywords to identify self-admitted technical debt in different sources?}}
\label{sec:rq2}

\begin{table}[b]
\caption{Top SATD keywords from different sources.}
\label{tb:keyword_diff_sources}
\begin{center}
\resizebox{0.8\columnwidth}{!}{
\def\arraystretch{1.2}
\begin{tabular}{cccc}
\hline
\textbf{Comment} & \textbf{Commit} & \textbf{Pull} & \textbf{Issue} \\
\cline{1-1}
\cline{2-2}
\cline{3-3}
\cline{4-4}
 \underline{hack} & typo & \underline{nit} & typo \\
 \underline{todo} & unused & typo & leak \\
 \underline{workaround} & unnecessary & unnecessary & flaky \\
 \underline{defer argument checking} & cleanup & redundant & unnecessary \\
 \underline{fixme} & simplify & simplify & \underline{performance} \\
 \underline{not needed} & leak & flaky & \underline{checkstyle} \\
 \underline{implement} & flaky & unused & \underline{spelling} \\
 \underline{this needs an extra} & redundant & \underline{confusing} & unused \\
 better & \underline{style} & cleanup & cleanup \\
 \underline{efficient} & \underline{polished} & better & \underline{coverage} \\
\hline
\end{tabular}}
\end{center}
\end{table}

\begin{figure}[htb]
  \centering
  \includegraphics[width=0.7\linewidth]{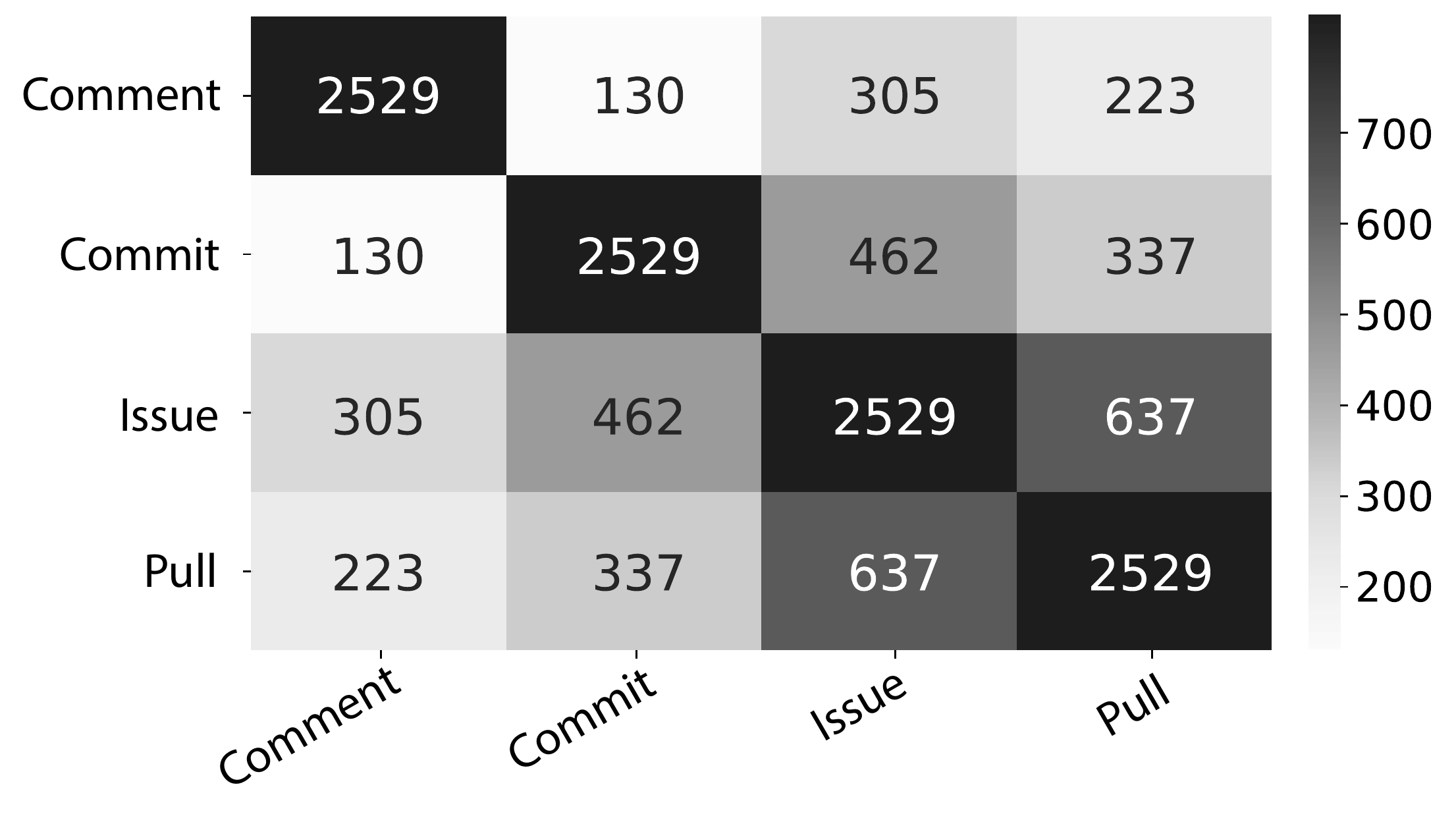}
  \caption{Number of shared keywords between different sources.}
  \label{f:keyword_correation}
\end{figure}

Using the method described in \cref{sec:keyword_extraction}, we summarize and present the top SATD keywords from four different sources (i.e., code comment, commit message, pull request, and issue) in \cref{tb:keyword_diff_sources}.
It is noted that the unique keywords (i.e., those that appear in only one source) are underlined.
Further, we calculate the average number of extracted SATD keywords for the different sources and utilize the top 10\% (i.e., 2529) of this average number of keywords to calculate the number of shared keywords between different sources. %
We choose the top 10\% similarly to our previous work, where we utilized the top 10\% of keywords to analyze relations between SATD in different sources \citep{li2022identifying}.
Our premise is that the more SATD keywords are shared between two sources, the more similar they are regarding SATD documentation. Consequently, we create a correlation matrix to show the number of shared SATD keywords between different sources (see \cref{f:keyword_correation}) to understand the similarities between SATD documentation in different sources.

In \cref{tb:keyword_diff_sources}, we can observe that source code comments have more unique SATD keywords compared with other sources.
This observation is consistent with the results in \cref{f:keyword_correation} that code comments have the least shared keywords with commit messages, pull requests, and issues (i.e., 130, 223, and 305 shared keywords respectively).
Moreover, commit messages have more shared SATD keywords with other sources compared to code comments.
Furthermore, issues have the greatest number of shared keywords with others, followed by pull requests.
The results indicate that issues and pull requests are the two most similar sources in terms of technical debt documentation, followed by commit messages, and finally by code comments.

\begin{table}[thb]
\caption{Top keywords for different types of SATD.}
\label{tb:keyword_diff_types}
\begin{center}
\resizebox{0.65\columnwidth}{!}{
\def\arraystretch{1.2}
\begin{tabular}{cc}
\hline
\textbf{Code/Design Debt} & \textbf{Documentation Debt} \\
\cline{1-1}
\cline{2-2}
unnecessary & typo \\
nit & spelling \\
leak & function needs documentation \\
unused & todo document \\
cleanup & missing license \\
simplify & document why \\
redundant & improve tutorial \\
performance & add some javdoc \\
checkstyle & add a comment \\
confusing & more documentation \\
\hline
\textbf{Test Debt} & \textbf{Requirement Debt} \\
\cline{1-1}
\cline{2-2}
flaky & not implemented \\
coverage & not thread-safe \\
flakiness & todo \\
todo test & work in progress \\
more tests & yet implemented \\
add tests & hasn't implemented \\
temporary test code & isn't thread safe \\
haven't tested & not safe \\
add a test & doesn't support \\
missing tests & isn't implemented \\
not tested & not supported \\
\hline
\end{tabular}}
\end{center}
\end{table}

In \cref{tb:keyword_diff_types}, we summarize the keywords for different types of SATD (i.e., code/design debt, documentation debt, test debt, and requirement debt).
We can notice that keywords for code/design debt largely overlap with the summarized top keywords in \cref{tb:keyword_diff_sources} because code/design debt is the most common type of SATD in different sources \citep{li2022self}.
We also note that some keywords could indicate more than one type of SATD.
For example, \emph{simplify} in \cref{tb:keyword_diff_types} indicates code/design debt, because it might refer to \emph{complex code} that is a kind of \emph{code debt}; but, it could also be used to indicate \emph{expensive test} which is a type of \emph{test debt}.

\begin{framed}
\noindent \textit{Issues and pull requests are the two most similar sources in terms of self-admitted technical debt documentation, followed by commit messages, and finally by code comments.}
\end{framed}

\subsection{(RQ3) \textit{How much and what types of self-admitted technical debt are documented in different sources?}}
\label{sec:rq3}

To answer this research question, we first train our proposed machine learning model with the best settings described in \cref{sec:rq1}.
Then, we use the trained machine learning model to classify the collected data from 103 projects (see \cref{sec:data_collection}) into four types of SATD, namely code/design debt, documentation debt, test debt, and requirement debt.

\begin{table}[thb]
\caption{Number and percentage of different types of SATD items identified from different sources.}
\label{tb:satd_num}
\begin{center}
\resizebox{\columnwidth}{!}{
\def\arraystretch{1.2}
\begin{tabular}{cccccccccccc}
\hline
\multirow{3}{*}{\textbf{Source}} & \multirow{3}{*}{\textbf{Total \#}} & \multicolumn{10}{c}{\textbf{Type of SATD}} \\
\cline{3-12}
 &  & \multicolumn{2}{c}{\textbf{Code/Design}} & \multicolumn{2}{c}{\textbf{Doc.}} & \multicolumn{2}{c}{\textbf{Test}} & \multicolumn{2}{c}{\textbf{Req.}} & \multicolumn{2}{c}{\textbf{All}} \\
\cline{3-4}
\cline{5-6}
\cline{7-8}
\cline{9-10}
\cline{11-12}
 & & \# & \% & \# & \% & \# & \% & \# & \% & \# & \% \\
\hline
Comment & 9,747,914 & \textbf{411,060} & 4.2 & 21,817 & 0.2 & 16,152 & 0.2 & \textbf{61,256} & \textbf{0.6} & \textbf{510,285} & 5.2 \\
Commit & 917,010 & 76,074 & 8.3 & 20,107 & \textbf{2.2} & 6,689 & 0.7 & 1,127 & 0.1 & 103,997 & 11.3 \\
Pull & 2,925,540 & 335,005 & \textbf{11.5} & \textbf{61,452} & 2.1 & \textbf{36,575} & \textbf{1.3} & 5,667 & 0.2 & 438,699 & \textbf{15.0} \\
Issue & 3,511,125 & 366,219 & 10.4 & 50,752 & 1.4 & 36,499 & 1.0 & 4,470 & 0.1 & 457,940 & 13.0 \\
\hline
\end{tabular}
}
\end{center}
\end{table}

\begin{table}[b]
\caption{Examples of different types of SATD.}
\label{tb:satd_examples}
\begin{center}
\resizebox{\columnwidth}{!}{
\def\arraystretch{1.2}
\begin{tabular}{cl}
\hline & \\[-2.5ex]
\textbf{Type of SATD} & \textbf{Example} \\
\hline
\textbf{Code/Design} & \makecell[l]{\textit{``Oh I didn't realize we got duplicated logic. We need to refactor this.''}\\ - [from Superset-pull-request-6831]} \\
 & \makecell[l]{\textit{``Need to add better handling for hz instance cleanup.''}\\ - [from Camel-jira-issue-10563]} \\
 & \makecell[l]{\textit{``Some new, friendlier APIs may be called for.''}\\ - [from Druid-github-issue-5940]} \\
\hline
\textbf{Documentation} & \makecell[l]{\textit{``Could you also please document the meaning of the various metrics''}\\ - [from Spark-pull-request-6905]} \\
 & \textit{``I think we should document this''} - [from Accumulo-jira-issue-1905] \\
 & \makecell[l]{\textit{``Currently, the api docs are missing from our website.''}\\ - [from Mxnet-github-issue-6648]} \\
\hline
\textbf{Test} & \makecell[l]{\textit{``It'd be good to add some usages of DurationGranularity to the query}\\ \textit{tests''} - [from Druid-github-issue-3994]} \\
 & \makecell[l]{\textit{``I did another cycle of review the unit tests, sorry I still not see value}\\ \textit{in denial-of-service tests?''} - [from Zookeeper-pull-request-689]} \\
 & \makecell[l]{\textit{``I would like to have at least a simple testcase around}\\ \textit{the UseV2WireProtocol feature''} - [from Bookkeeper-github-issue-272]} \\
\hline
\textbf{Requirement} &  \makecell[l]{\textit{``TODO: add a dynamic context in front of every selector with a}\\ \textit{traversal''} - [from Heron-code-comment]} \\
 & \makecell[l]{\textit{``Remaining todo list for SQL parse module...''}\\ - [from Pinot-github-issue-2505]} \\
 & \makecell[l]{\textit{``Union is not supported yet. But i might be adding that capability}\\ \textit{quite soon.''} - [from Samza-pull-request-295]} \\
\hline
\end{tabular}
}
\end{center}
\end{table}

\cref{tb:satd_num} presents the number and percentage of the four types of SATD identified from different sources.
We can observe that most SATD items are identified from source code comments, followed closely by issues and pull requests (i.e., $510,285$, $457,940$, and $438,699$ respectively).
Commit messages have the least SATD items (i.e., $103,997$), corresponding to about one-fifth of SATD identified from code comments.
In contrast to the number of SATD items, we can notice that code comments have the lowest percentage of SATD ($5.2\%$), while pull requests have the highest percentage of SATD ($15.0\%$).
Moreover, regarding the different types of SATD, while code comments have the lowest percentage of code/design debt ($4.2\%$), they contain the largest amount of code/design debt ($411,060$) compared to other sources.
A significant amount of code/design debt is also identified from pull requests and issues ($335,005$ and $366,219$ respectively).
Compared to code/design debt, other types of SATD have significantly fewer SATD items.
Documentation debt is relatively evenly admitted in different sources.
The number of documentation debt items ranges from $21,817$ to $61,452$ in the four sources, while most of them are documented in pull requests and issues ($61,452$ and $50,752$).
Furthermore, we can see that pull requests and issues contain more test debt ($36,575$ and $36,499$ respectively) compared to the other two sources ($16,152$ and $6,689$).
Lastly, we notice the vast majority of requirement debt is documented in code comments ($61,256$) compared to the other three sources ($5,667$, $4,470$, and $1,127$).
To provide some insight into what the different types of SATD look like, we provide some identified representative examples from each type in \cref{tb:satd_examples}.
We discuss four examples here, based on the definitions presented in \cref{tb:statistics_type_projects}.
The first example of code/design debt (i.e., \textit{I didn't realize we got duplicated logic. We need to refactor this}) is \textit{duplicated code} which is a typical manifestation of code/design debt.
Moreover, the example of documentation debt (i.e., \textit{Could you also please document the meaning of the various metrics}) indicates \emph{low-quality documentation}.
Next, the first example of test debt (i.e., \textit{It'd be good to add some usages of DurationGranularity to the query tests}) refers to \emph{low coverage} which leads to test debt.
Lastly, \textit{union is not supported yet} indicates \emph{partially implemented requirements} which indicates requirement debt.

\begin{table}[htb]
\caption{Details of contribution flows.}
\label{tb:units}
\begin{center}
\resizebox{0.85\columnwidth}{!}{
\def\arraystretch{1.2}
\begin{tabular}{rccc}
\hline
\textbf{Contribution Flow} & \textbf{Abbr.} & \textbf{\#} & \textbf{\%} \\
\hline
{\textbf{Issue} \textrightarrow~\textit{Pull(s)} \textrightarrow~Commit(s) \textrightarrow~Comment(s)} & IPCC & 81,940 & 8.5 \\
{\textbf{Issue} \textrightarrow~Commit(s) \textrightarrow~Comment(s)} & ICC & 182,406 & 18.9 \\
{\textit{Pull} \textrightarrow~Commit(s) \textrightarrow~Comment(s)} & PCC & 109,621 & 11.3 \\
{Commit \textrightarrow~Comment(s)} & CC & 593,015 & 61.3 \\
\hline
\end{tabular}
}
\end{center}
\end{table}

Based on the links built between sources (see \cref{sec:link_sources,}) and the workflow (see \cref{f:flow}), we summarize four types of contribution flows in \cref{tb:units} (note the abbreviation of each contribution flow).
As can be seen, the most common way to contribute is directly pushing commits to change source code (61.3\%), which is followed by \emph{ICC} (18.9\%).
Furthermore, \emph{PCC} and \emph{IPCC} are the least common contribution flows (11.3\% and 8.5\% respectively).
To help readers gain a better understanding of the contribution flows, we show an example of the contribution flow \emph{IPCC}:

\begin{enumerate}
    \item Developers first created an issue\footnote{\url{https://github.com/apache/tvm/issues/2351}} (\emph{\#2351}) to support quantized models.
    This work consists of four tasks, as numbered below:
    
    \begin{quote}
        \textit{``[RFC][Quantization] Support quantized models from TensorflowLite...
        \\1. Support TFLite FP32 Relay frontend. PR: \textbf{\#2365}
        \\2. Support TFLite INT8 Relay frontend
        \\3. Extend the attribute of the convolution and related ops to support quantization
        \\4. Auto-TVM on ARM CPU can work with INT8...''} - [from Tvm-github-issue-2351]
    \end{quote}
    
    \item Subsequently, they created a pull request\footnote{\url{https://github.com/apache/tvm/pull/2365}} (\emph{\#2365}) to work on the first task.
    The number of the pull request (\emph{\#2365}) is then added in the issue description to link these two sources.
    Meanwhile, the related issue number (\emph{\#2351}) is included in the pull request description:
    
    \begin{quote}
        \textit{``[TFLite] Support TFLite FP32 Relay frontend.
        \\This is the first PR of \textbf{\#2351} to support importing exist quantized int8 TFLite model. The base version of Tensorflow / TFLite is 1.12.''} - [from Tvm-pull-request-2365]
    \end{quote}
    
    \item After discussing solutions and code review, a commit\footnote{\url{https://github.com/apache/tvm/commit/10df78a}} with code changes was merged into the master branch. 
    In the commit message, the related pull request (\emph{\#2365}) was mentioned:
    
    \begin{quote}
        \textit{``[TFLite] Support TFLite FP32 Relay frontend. (\textbf{\#2365})
        \\- Support TFLite FP32 Relay frontend.
        \\- Fix lint issue
        \\- Remove unnecessary variables and packages...''} - [from Tvm-commit-10df78a]
    \end{quote}
    
    \item Finally, code comments were added to the repository when merging the commit:
    
    \begin{quote}
        \textit{``\# add more if we need target shapes in future''} - [from Tvm-code-comment-10df78a]
    \end{quote}
\end{enumerate}

\begin{figure}[thb]
  \centering
  \includegraphics[width=0.9\linewidth]{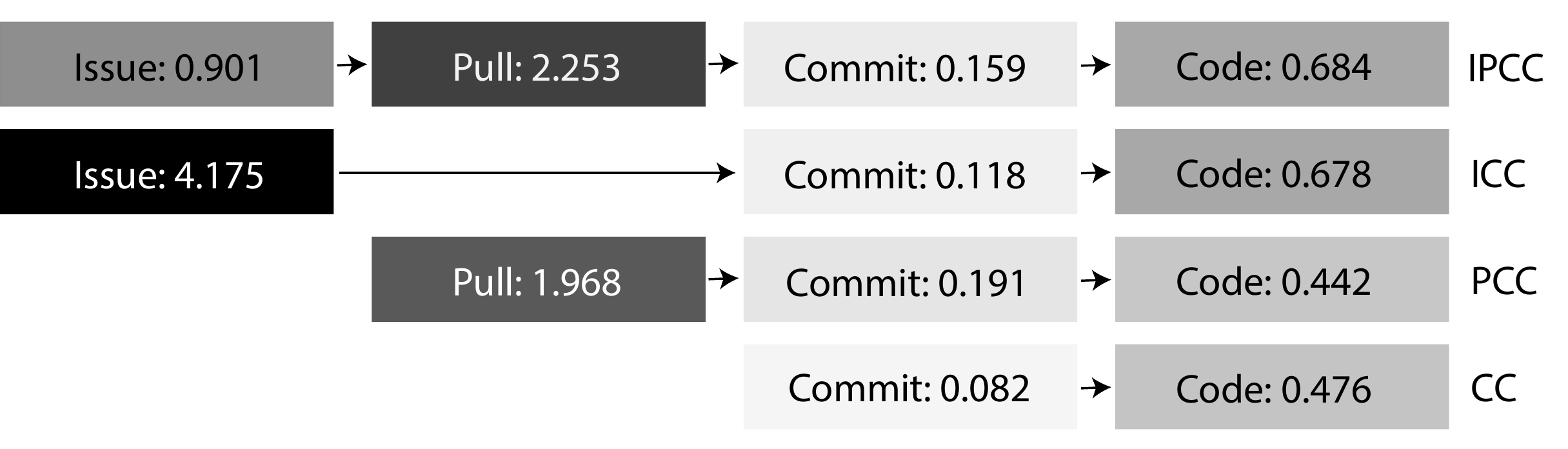}
  \caption{Average number of SATD items in different contribution flows.}
  \label{f:satd_num_flow}
\end{figure}

The four flows for all analyzed data as listed in \cref{tb:units}, are independent of the existence of SATD. 
Subsequently, we analyze and present the average number of SATD items in different sources with regard to the four types of contribution flows.
The average number of SATD items per source is illustrated in \cref{f:satd_num_flow}, again for all analyzed data.
To obtain this average number for each contribution flow, we divide the number of SATD items in a specific source by the number of all SATD items in this contribution flow.
For example, consider that five SATD items are identified in code comments in ten \emph{IPCC} flows.
The average number of SATD items in code comments in \emph{IPCC} is thus $\frac{5}{10} = 0.5$.
We notice that for contribution flows \emph{IPCC}, \emph{ICC}, and \emph{PCC}, the sum of the averages is more than two. This is because issues and pull requests are composed of multiple issue sections and pull request sections, and there can be more than one related pull request for each issue.

In comparison with issues and pull requests, there is less than one SATD on average identified from commit messages or code comments for all contribution flows. 
It is noted that even though the average number of SATD items in commits and code comments is low because there is a huge amount of the contribution flow \emph{CC}, the number of SATD in these two sources is still comparable to the other two sources (see \cref{tb:satd_num}).
More specifically, for contribution flows \emph{IPCC} and \emph{ICC} that both start with an issue, they have more technical debt admitted in code comments on average compared to \emph{PCC} and \emph{CC} ($0.684$ and $0.678$ versus $0.442$ and $0.476$).
Moreover, comparing \emph{IPCC} and \emph{ICC}, when developers do not use pull requests, significantly more SATD is documented in issues ($4.175$ versus $0.901$).
Furthermore, we also observe that when developers choose to use pull requests (see \emph{IPCC} and \emph{PCC} in \cref{f:satd_num_flow}), more technical debt is admitted in commit messages ($0.159$ and $0.191$ versus $0.118$ and $0.082$).

\begin{framed}
\noindent \textit{SATD is evenly spread among sources (i.e., source code comment, commit message, pull request, and issue tracker).
There are more than two SATD items identified on average for contribution flows that use issues or pull requests.
When developers do not use pull requests, significantly more SATD is documented in issues (4.175 versus 0.901).}
\end{framed}

\subsection{(RQ4) \textit{What are the relations between self-admitted technical debt in different sources?}}
\label{sec:rq4}

\begin{figure}[t]
  \centering
  \includegraphics[width=0.85\linewidth]{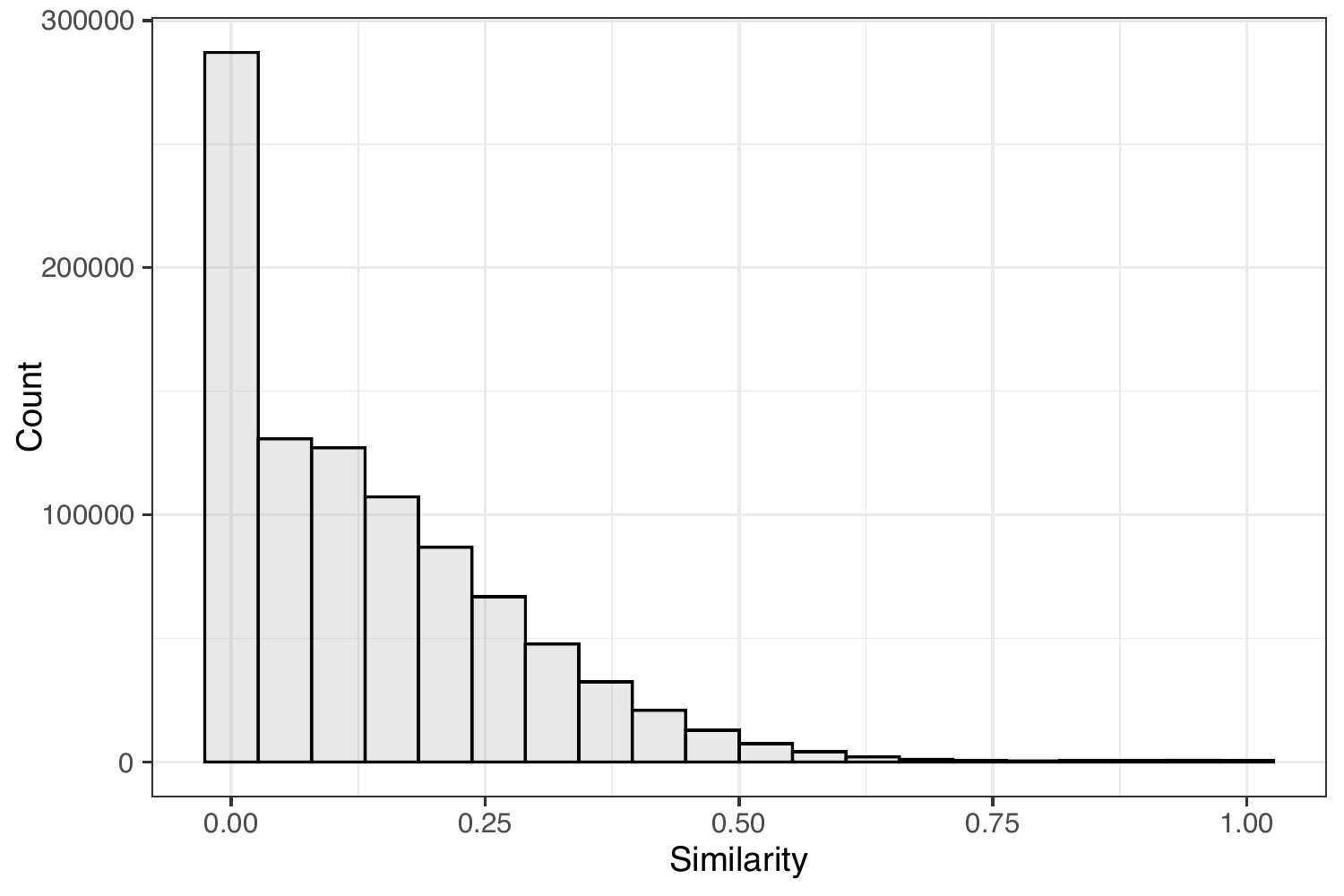}
  \caption{Histogram of similarity scores.}
  \label{f:similarity_histogram}
\end{figure}

To understand the relations between SATD in different sources, as described in \cref{sec:satd_similarity}, we first use the cosine similarity to determine the similarity between SATD items.
When answering RQ3 (see \cref{sec:rq3}), we observed that SATD in different contribution flows typically refer to different technical debt, even if their textual information is similar.
For example, there are two commits about fixing typos with the messages of \textit{Fix typo - [from Camel-6b5f64a]} and \textit{Typo - [from Camel-a41323c]}.
The similarity between these two commits is high, but they are referring to different typos.
Therefore, we only analyze the similarity between SATD in the same contribution flows to avoid this situation.
The analysis results in the distribution of similarity score are illustrated in \cref{f:similarity_histogram}.
The results indicate that the average similarity score is $0.135$ with a standard deviation of $0.142$, which entails an uneven distribution of the similarity score.

\begin{figure}[b]
  \centering
  \includegraphics[width=0.85\linewidth]{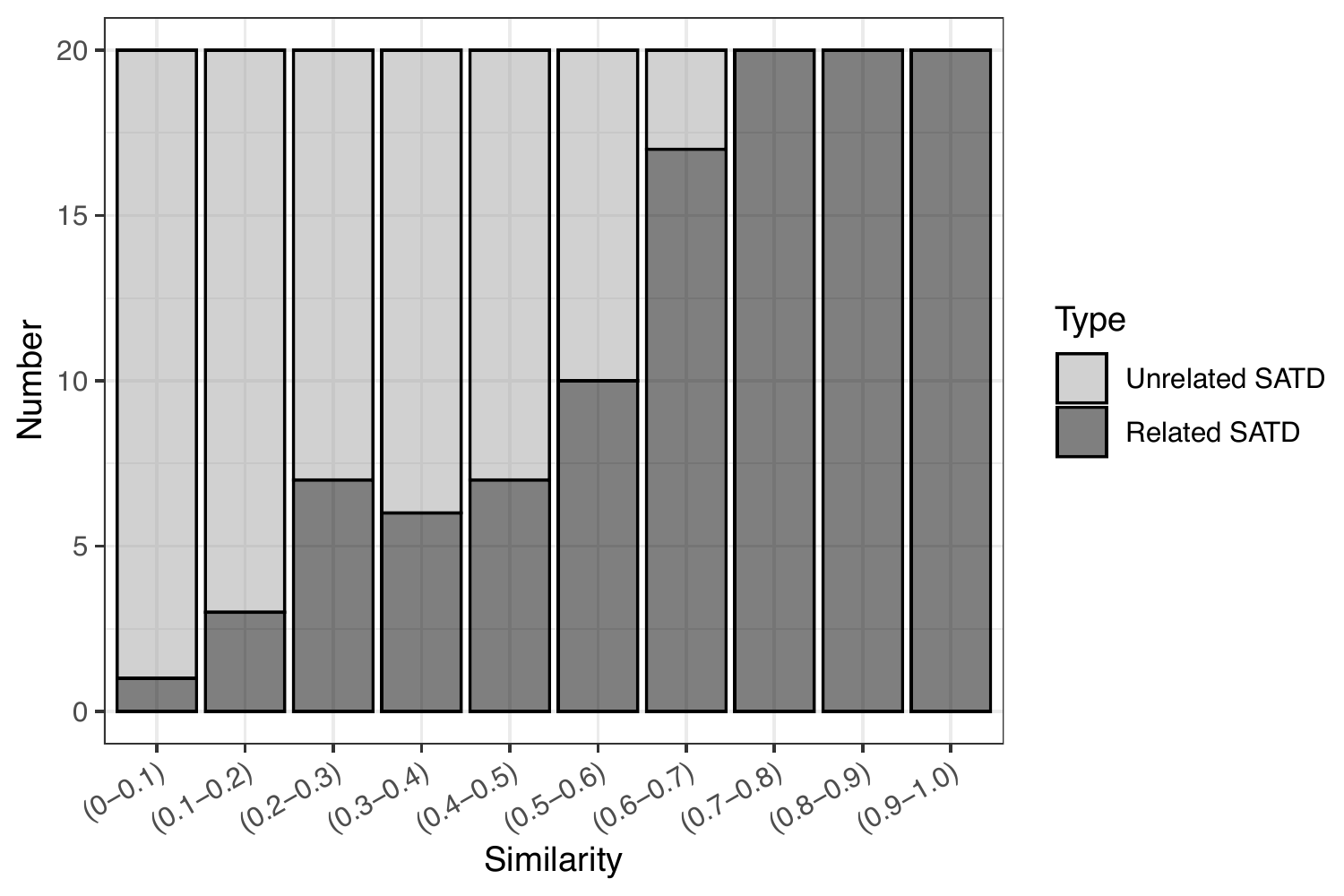}
  \caption{Percentage of pairs discussing related or unrelated SATD against cosine similarity scores.}
  \label{f:same_satd_differnet_source_percentage}
\end{figure}

To distinguish between unrelated SATD items and related SATD items, we used the stratified sampling method to get 10 groups of samples (each group containing 20 pairs of items) with similarity scores between 0 and 0.1, 0.1 and 0.2,... 0.9 and 1.0.
Then the first author and second author independently manually analyzed the samples and classified them as related SATD or unrelated SATD.
After that, we evaluated the level of agreement between the classifications of the two authors using Cohen's kappa coefficient \citep{landis1977measurement} to measure the inter-rater reliability.
The obtained Cohen's kappa coefficient is $+0.81$, which is an `almost perfect' agreement according to the work of \cite{landis1977measurement}.%
The results of the classification are presented in \cref{f:same_satd_differnet_source_percentage}.
As can be seen, when the similarity score is between $0.4$ and $0.5$, only 7 out of 20 examples are related SATD.
When the similarity score is between $0.5$ and $0.6$, 10 out of 20 examples are referring to the related SATD.
Therefore, we consider two SATD items to be related when the similarity score is greater than $0.5$; we discuss this further (and potential threats to validity) in Section \ref{sec:discussion}.

\begin{table}[htb]
\caption{Number of the related SATD items documented in pairs of sources.}
\label{tb:satd_relation_num}
\begin{center}
\resizebox{0.65\columnwidth}{!}{
\def\arraystretch{1.2}
\begin{tabular}{rlcc}
\hline
\multicolumn{2}{c}{\textbf{Pair of Sources}} & \textbf{Number} & \textbf{Total} \\
\hline
\multirow{3}{*}{Code Comment $\leftrightarrow$} & Commit & 482 & \multirow{3}{*}{3,747} \\
 & Pull Request & 989 &  \\
 & Issue & \textbf{2,276} \\
\hline
\multirow{3}{*}{Commit $\leftrightarrow$} & Code Comment & 482 & \multirow{3}{*}{3,564} \\
 & Pull Request & 1,746 & \\
 & Issue & 1,336 \\
\hline
\multirow{3}{*}{Pull Request $\leftrightarrow$} & Code Comment & 989 & \multirow{3}{*}{3,829} \\
 & Commit & 1,746 \\
 & Issue & 1,094 \\
\hline
\multirow{3}{*}{Issue $\leftrightarrow$} & Code Comment & \textbf{2,276} & \multirow{3}{*}{\textbf{4,706}} \\
 & Commit & 1,336 \\
 & Pull Request & 1,094 \\
\hline
\end{tabular}
}
\end{center}
\end{table}

\cref{tb:satd_relation_num} shows how many of the related SATD items are documented in two different sources.
As we can see, the most common combination of sources for admitting related technical debt items is issues and code comments (2,276), followed by pull requests and commits (1,746).
The least common way to document related technical debt is in code comments and commits (482).
Furthermore, comparing the number of SATD items documented in a combination of one specific source and the other three sources, we can observe that the combination of issues and others has the greatest number of SATD items (4,706).
However, the differences between this combination and other combinations (3,829, 3,747, and 3,564) are not significant.
This indicates that the numbers of related SATD items in different sources are comparable.

\begin{figure}[htb]
  \centering
  \includegraphics[width=0.85\linewidth]{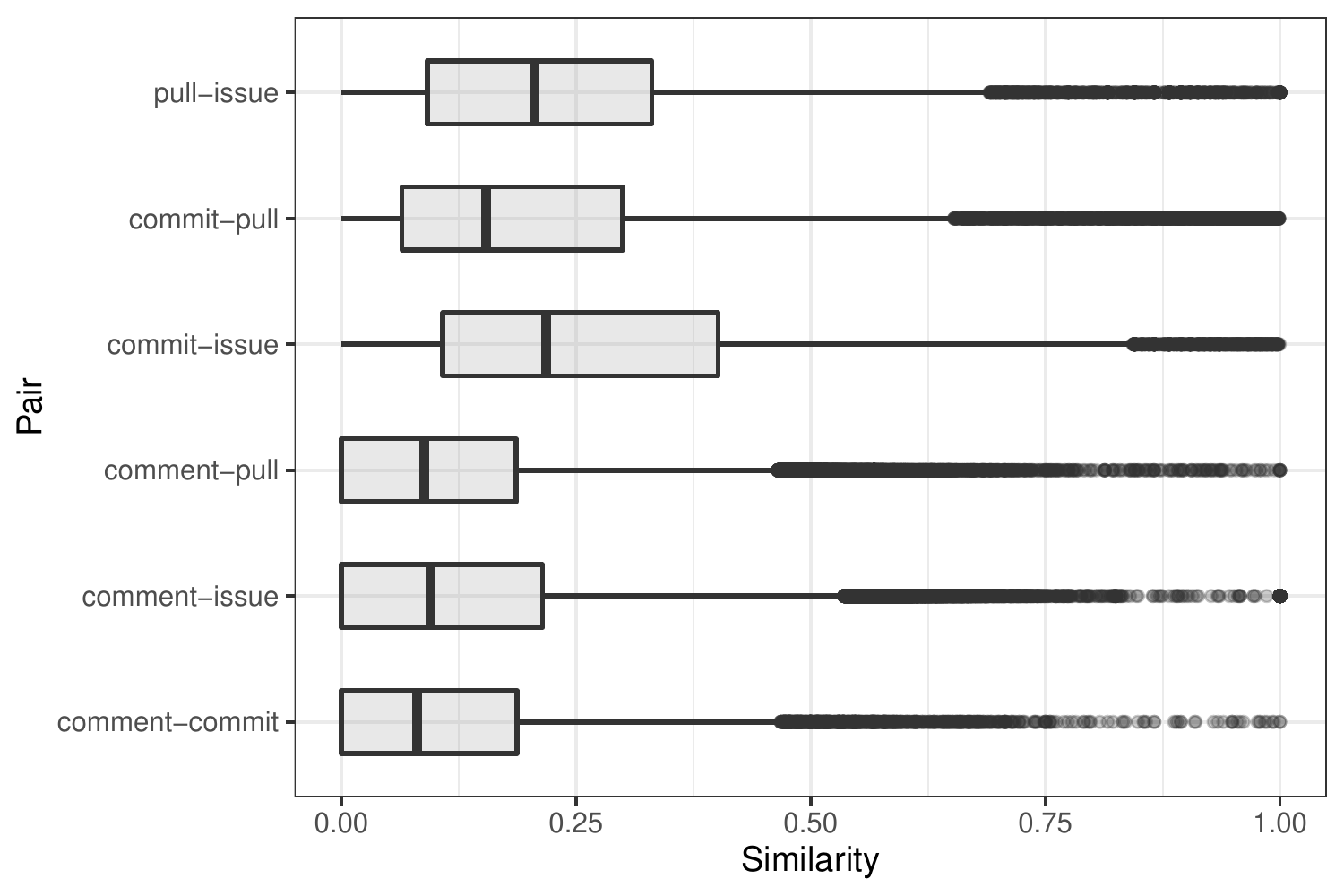}
  \caption{Cosine similarity distribution in pairs of sources.}
  \label{f:similarity_between_sources}
\end{figure}

Moreover, \cref{f:similarity_between_sources} presents the distributions of cosine similarity in pairs of sources.
With a visual inspection, we see that the median similarity between SATD in code comments and other sources is lower than the other combinations.
Furthermore, the pairs of commit-issue and pull-issue show a slightly higher median similarity than the pair of commit-pull.

\begin{figure}[htb]
  \centering
  \includegraphics[width=0.7\linewidth]{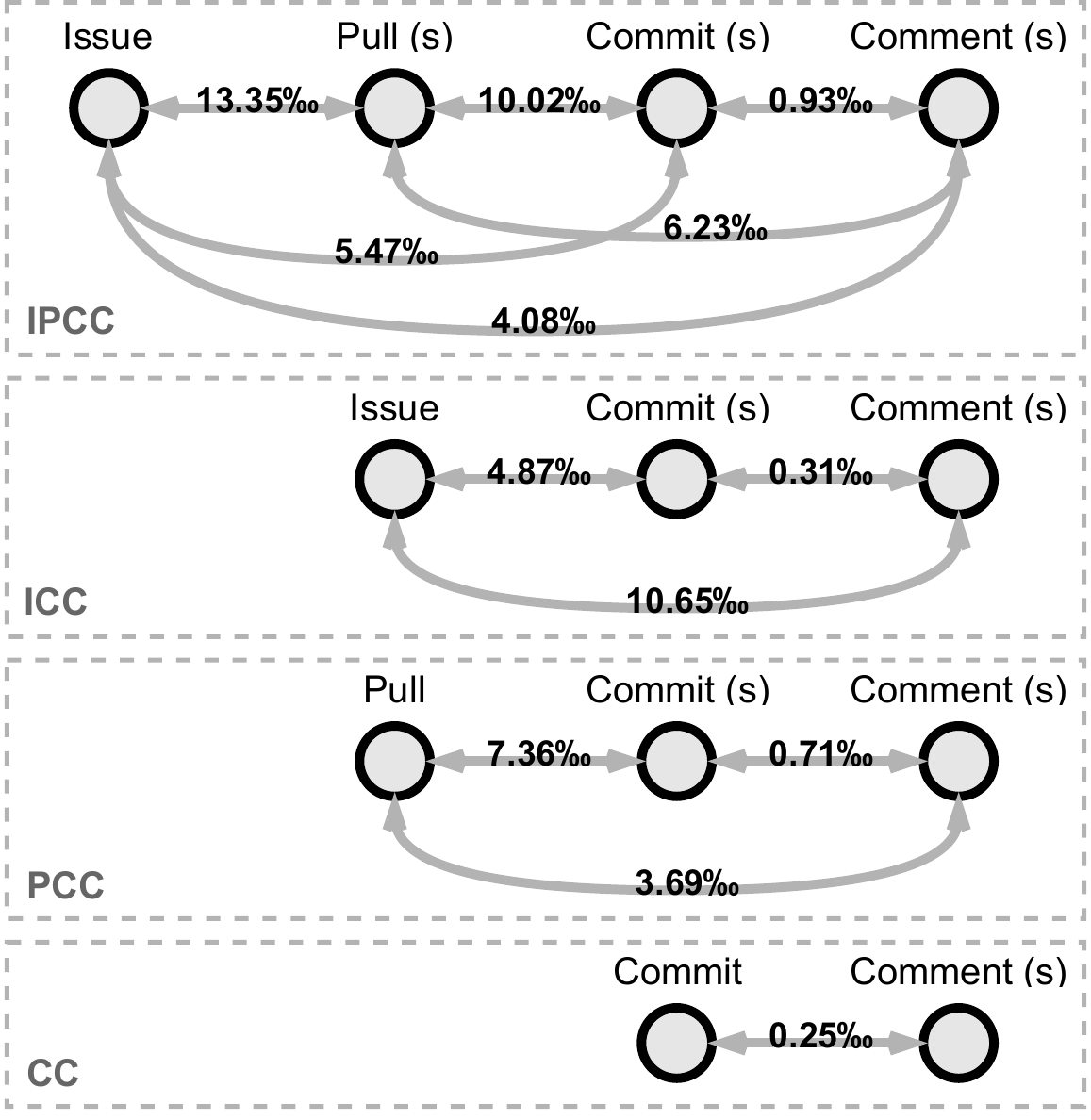}
  \caption{Percentage of comments referring to the related SATD regarding different contribution flows.}
  \label{f:satd_relations}
\end{figure}

Additionally, to explore the connections between related SATD and contribution flows, we calculate the permillage of related SATD in pairs of sources in different contribution flows (see \cref{f:satd_relations}).
In the contribution flow \emph{IPCC}, we can observe that developers tend to document related SATD in the adjacent sources.
For example, considering the technical debt admitted in an issue, the probabilities of related SATD documented in pulls, commits, and code comments are 13.35\textperthousand, 5.47\textperthousand, and 4.08\textperthousand~respectively.
Furthermore, we notice that developers document related SATD in issues and code comments more frequently in \emph{ICC} than in \emph{IPCC}.
Furthermore, there are fewer chances that related SATD is documented in different pairs of sources in \emph{PCC} in comparison with \emph{IPCC}.

Finally, we conducted a qualitative analysis according to the \emph{Constant Comparative} method \citep{glaser1965constant} on 200 examples of related SATD items to investigate how developers take advantage of SATD in multiple sources.
Specifically, the first and second authors coded the related SATD items separately.
Then they compared the results and reached an agreement on the following four types of relations:

\begin{itemize}
    \item \textbf{Documenting existing SATD in additional sources.}
    We found that developers document already existing SATD in other sources for two different reasons.
    As shown in \cref{f:flow}, when developers identify technical debt and discuss it in issues or pull requests, if they choose not to fix it immediately, they could document it in code comments or commit messages, as a reminder to repay it in the future.
    For example, a developer agreed to improve functionality, but not immediately.
    They then commented in the pull request:

    \begin{displayquote}
        \textit{``...to improve the read throughput, creating new watcher bit and adding it to the BitHashSet has its own lock to minimize the lock scope. \textbf{I'll add some comments here}.''} - [from Zookeeper-pull-590]
    \end{displayquote}
    
    Subsequently, they created a code comment to point out the issue that needs to be resolved:
    
    \begin{displayquote}
        \textit{``// Need readLock to exclusively lock with removeWatcher, otherwise we may add a dead watch whose connection was just closed. Creating new watcher bit and adding it to the BitHashSet has it's own lock to minimize the write lock scope.''} - [from Zookeeper-code-comment]
    \end{displayquote}
    
    A second case arises when developers report technical debt in issues and decide to solve it with pull requests; they often create a new pull request using the same title or description as the issue to describe the existing SATD. 
    For example, a developer created an issue to solve a legacy code problem:
    
    \begin{displayquote}
        \textit{``Cleanup the legacy cluster mode.''} - [from Tajo-issue-1482]
    \end{displayquote}
    
    After discussion, developers chose to create a pull request to pay back the debt:
    
    \begin{displayquote}
        \textit{``TAJO-1482: Cleanup the legacy cluster mode.''} - [from Tajo-pull-484]
    \end{displayquote}

    \item \textbf{Discussing the solution of SATD in other sources.}
    When technical debt is reported in issues, developers may choose to create a pull request to discuss detailed solutions for it (see \cref{f:flow}).
    For example, a developer reported a problem with mixing public and private headers by creating an issue:
    
    \begin{displayquote}
        \textit{``Some public headers include private headers. Some public headers include items that do not need to be included.''} - [from Geode-issue-4151]
    \end{displayquote}
    
    After that, they described the details of this technical debt and discussed the solutions in a pull request:
    
    \begin{displayquote}
        \textit{``I found that early on we had mixed up the include paths in the CMake project so we were able to include private headers from the public headers. This will cause anyone trying to build a client to have a frustrating time since public won't be able to find private headers...''} - [from Geode-pull-173]
    \end{displayquote}

    \item \textbf{Documenting the repayment of SATD in other sources.}
    When SATD is paid back, this repayment is sometimes documented in other sources.
    As we can see in \cref{f:flow}, when the SATD is solved after discussing it inside issues or pull requests, developers could document its repayment in commit messages or code comments.
    For example, a software engineer found that error messages are too general and reported it in an issue:
    
    \begin{displayquote}
        \textit{``To make troubleshooting easier I think that a more fine grained error handling could provide the user with a better view of what the underlying error really is.''} - [from Camel-issue-9549]
    \end{displayquote}
    
    When the error messages were improved, the engineer reported the SATD repayment in the commit message:
    
    \begin{displayquote}
        \textit{``CAMEL-9549 - Improvement of error messages when compiling the schema.''} - [from Camel-commit-dc3bb68]
    \end{displayquote}
    
    Additionally, it is also usual to document SATD repayment in source code comments.
    For example, a software engineer reported a code duplication problem by creating a Jira issue ticket:
    
    \begin{displayquote}
        \textit{``...a lot of functionality is shared between Followers and Observers. To avoid copying code, it makes sense to push the common code into a parent Peer class and specialise it for Followers and Observers.''} - [from Zookeeper-issue-549]
    \end{displayquote}
    
    When this technical debt was solved, the engineer added an explanation in the code comments for this SATD repayment:
    
    \begin{displayquote}
        \textit{``// This class is the superclass of two of the three main actors in a ZK ensemble: Followers and Observers. Both Followers and Observers share a good deal of code which is moved into Peer to avoid duplication.''} - [from Zookeeper-code-comment]
    \end{displayquote}
    
    \item \textbf{Paying back documentation debt in code comments.}
    This is a special case of the previous one. Because code comments are a kind of documentation, some documentation debt can be paid back by adding comments or Javadoc in source code comments.
    When documentation debt is reported in issues, developers might pay back the debt directly by writing code comments (see \cref{f:flow}).
    For example, a developer found that documentation is incomplete:

    \begin{displayquote}
        \textit{``If the assumption is that both the buffers should be of same length, please document it.''} - [from Pinot-pull-2983]
    \end{displayquote}
    
    Subsequently, they updated the source code comments to solve this debt:
    
    \begin{displayquote}
        \textit{``// NOTE: we don't check whether the array is null or the length of the array for performance concern. All the dimension buffers should have the same length.''} - [from Pinot-code-comment]
    \end{displayquote}
\end{itemize}

\begin{framed}
\noindent \textit{The numbers of related SATD items in different sources are comparable, while code comments and issues have the greatest number of related SATD items compared to other combinations.
There are four types of relations between SATD in different sources: 1) documenting existing SATD repeatedly; 2) discussing the solution of SATD; 3) documenting the repayment of SATD; 4) repaying documentation debt in code comments.}
\end{framed}

\section{Discussion}
\label{sec:discussion}

\subsection{Automatic Identification of Different SATD Types in Multiple Sources}

In recent years, a great number of studies explored various SATD identification approaches \citep{SIERRA201970}.
However, there has been very limited work on identifying different types of SATD (i.e., design, requirement, documentation, and test debt).
To the best of our knowledge, there are only two works \citep{da2017using,chen2021multiclass} focusing on detecting different types of SATD: one identified design debt and requirement debt using a maximum entropy classifier \citep{da2017using}, while the other utilized several machine learning methods to identify design, requirement, and defect debt \citep{chen2021multiclass}.
Test and documentation debt were ignored in these two works, while both of them identified SATD only from source code comments.
In this study, we not only identify SATD from four sources but also identify four different types of SATD (i.e., code/design, requirement, documentation, and test debt).
In comparison with the two aforementioned studies \citep{da2017using,chen2021multiclass}, our average F1-score for identifying different types of SATD from code comments is superior (0.667 versus 0.512 and 0.558).
Therefore, we encourage practitioners to design and implement SATD tracking systems based on our multi-type SATD identifier to facilitate the repayment of SATD in different sources and help developers get better insights into SATD.
However, we still notice that the machine learning model is struggling to identify test and requirement debt (see \cref{tb:f1_all_methods}).
\textbf{Thus, we suggest that researchers further improve the identification performance by enriching the datasets or optimizing identification approaches.}

Moreover, researchers could investigate how to prevent missing SATD documentation based on our SATD identifier.
For example, if developers in a team are advised to only document SATD in code comments or issues, but the identifier finds instead that SATD is only documented in commit messages, then action must be taken to motivate and train developers to use the recommended sources for SATD documentation.
Furthermore, researchers could develop new approaches based on our identification approach to detect erroneous SATD or SATD that bears a high risk.

Additionally, because our approach is able to extract SATD keywords, practitioners could design guidelines or even a standard for documenting SATD and utilize our approach to examine whether keywords are used well or misused for SATD documentation.
Meanwhile, according to the results demonstrated in \cref{sec:rq4}, we found that technical debt is documented in different ways in software artifacts.
Some of the SATD items are about repayment, while others concern reporting or discussing of the existing SATD items.
To further investigate this phenomenon, we took a sample of 100 SATD items from pull request sections and commit messages, and analyzed the number of SATD items regarding reporting existing SATD and repaying SATD.
The results show that, in pull requests, 66\% of SATD items are about reporting existing SATD, 8\% of SATD items are about repaying existing SATD, while 13\% are unclear and require contextual information to determine.
Furthermore, in commit messages, 2\% report existing SATD, 88\% resolve existing SATD, and 10\% are unclear.
However, we currently lack tools that can automatically differentiate between the description and repayment of SATD.
This would offer two advantages.
First, practitioners could use this information to manage their technical debt.
For example, as shown in \cref{sec:rq4}, SATD repayment is sometimes documented in source code comments.
When developers want to resolve SATD in code comments, they need to check whether it concerns SATD introduction or repayment (the latter obviously does not need to be resolved).
If this was automated, developers could easily get a list of SATD items by filtering out SATD repayment.
Second, researchers could use this information to better study the nature of SATD.
For example, they could easily calculate how much SATD is introduced or paid back.
\textbf{We thus propose that researchers work on approaches to automatically differentiate between reporting and repaying SATD.}

Finally, in this work, we observed that some developers prefer to admit technical debt in code comments to be addressed, while some tend to document technical debt in issues or other sources.
Our results actually indicate that SATD is evenly spread in different sources (see \cref{sec:rq3}).
However, there are currently no tools that provide assistance in managing SATD across different sources.
Our proposed approach (MT-Text-CNN) is an initial step in this direction as it supports \textit{identifying} and \textit{relating} SATD in distinct sources. \textbf{We further advise researchers to investigate SATD visualization, prioritization, and repayment techniques across different sources, 
based on our SATD identification approach.}

\subsection{Self-Admitting Technical Debt in Different Sources}
\label{sec:diss_relations}

In \cref{sec:rq2,sec:rq3}, we summarized and presented the characteristics, keywords, and relations of SATD in different sources.
Observing \cref{tb:satd_num}, we found that although source code comments contain more SATD items (510,285) than other sources (457,940, 438,699, and 103,997), overall the SATD identified in all sources is comparable.
Since the majority of related work has investigated SATD in code comments \citep{SIERRA201970}, this finding indicates that the other data sources remain untapped. Thus, \textbf{considering the significant amount of SATD identified in sources other than source code (i.e., issue trackers, pull requests, and commit messages), we advise researchers to focus more on SATD in these sources.}
Moreover, according to the previous work \citep{guo2021far}, keyword-based approaches achieved very good performance when identifying SATD or non-SATD from code comments.
Practitioners could utilize the keywords summarized in this work to build lightweight keyword-based SATD detectors to identify different types of SATD from different sources and evaluate their predictive performance.
Furthermore, summarized keywords could be used for training purposes to give students or practitioners a better understanding of SATD.
Additionally, researchers could also evaluate the precision of the extracted keywords.
In \cref{tb:keyword_diff_types}, we found SATD keywords are different in different sources.
We suggest that researchers investigate whether it is a good practice to document SATD using different keywords in different sources.

In this work, we studied the relations between SATD in different sources by a) analyzing the number of shared SATD keywords between different sources (see \cref{f:keyword_correation}); and b) calculating the average cosine similarity score between SATD items from different sources (see \cref{f:similarity_between_sources}).
As we can see in these two figures, the relations between code comments and other sources are the weakest, followed by the relations between commits and pull requests.
Moreover, both figures indicate that the relations between issues and pull requests or commits are the strongest.
This could be caused by the nature of different sources: developers typically use issues and pull requests to solve problems, and then document the repayment of problems in commits (see \cref{f:flow}).
Additionally, our findings show that the related SATD documented in issues and code comments is the most common among all combinations (see \cref{tb:satd_relation_num}).
However, neither this nor the other relations have been investigated in previous work.
There was only one study that utilizes the relation between code comments and commit messages to study SATD repayment \citep{zampetti2018self}; all other relations (such as issues and code comments, see \cref{sec:rq4}) have not been investigated yet.
By leveraging these relations, researchers could better understand and improve SATD management activities.
For example, researchers could analyze the relation between SATD solution discussion and SATD documentation to analyze the impact of SATD, because the more significant the SATD, the more discussion it takes to resolve.
Considering the advantages of SATD relations, we suggest that \textbf{researchers link the SATD between different sources and make use of these links to further investigate SATD and support SATD management.}

Furthermore, when determining the threshold for the related SATD or unrelated SATD, we noticed that in some cases there are relations between two SATD items even if  the cosine similarity score is low (see \cref{f:same_satd_differnet_source_percentage}).
For example, developers discussed improving the logic of the code in a pull request:

\begin{displayquote}
    \textit{``...added this logic to make it easier for the FE (you can see it in the `create` logic already), by not requiring us to stringify our json beforehand, which I'm fine with. Do you see it as being an issue in the long run?''} - [from Superset-pull-11770]
\end{displayquote}

Then they chose not to solve it immediately, and reported that the logic needs to be improved in a code comment:

\begin{displayquote}
    \textit{``// Need better logic for this''} - [from Superset-code-comment]
\end{displayquote}

In this case, the cosine similarity of these two SATD items is only 0.22 (this is below our threshold of 0.5), while they are still referring to the related SATD.
\textbf{Therefore, we suggest researchers improve the SATD relation analysis algorithm to reduce false negative cases.}

Finally, there are also limits to the calculation of similarity between SATD in different contribution flows, because textual information is not sufficient to determine the relations of SATD in many cases.
For example, developers reported a typo in a pull request:

\begin{displayquote}
    \textit{``yes, agreed, it's a typo...''} - [from Drill-pull-602]
\end{displayquote}

However, it is not clear if this is fixed and where, as there are several commit messages documenting typo fixes, e.g.:

\begin{displayquote}
    \textit{``Fix typo''} - [from Drill-commit-c77f13c]
\end{displayquote}

In this situation, it is not possible to determine whether these SATD items refer to the same SATD item by only relying on textural information.
\textbf{Hence, researchers need to take other information (e.g., creation time, author, and code changes) into consideration to improve the SATD relation analysis.}

\section{Threats to Validity}
\label{sec:validity}

\subsection{Construct validity}

Threats to construct validity concern to what extent the operational measures represent what is investigated in accordance with the research questions. 
In this study, our main goal is to automatically identify SATD from different sources.
Because the used datasets are imbalanced (less than 15\% of the data is classified as SATD), simply measuring the proportion of correct predictions (both true positives and true negatives) among the total number of cases %
could be biased.
For example, assuming we have 15 SATD items out of 100 items, if all items are classified as non-SATD items, the classifier achieves an \emph{accuracy} of 85\%.
However, in this case, no SATD item is found by the identifier.
In another case, if the classifier correctly identifies 10 SATD items and 70 non-SATD items, the \emph{accuracy} of the predicted result is 80\%.
This case seems worse than the first one while it actually performs better in terms of SATD identification.
To eliminate this bias, we chose to use the same metric (i.e., \emph{F1-score}) as the previous study \citep{da2017using,ren2019neural,li2022identifying} as the \emph{F1-score} is a harmonic mean of the precision and recall.
Using the \emph{F1-score} as the metric, the measurement for the first and second cases are 0 and 0.5 respectively, making them a much better fit in evaluating the performance of classifiers.

Moreover, a possible threat to construct validity comes from the method of extracting SATD keywords to present the most informative keywords from different sources.
If the extracting method is inaccurate or unstable, the results could be erroneous.
To (partially) mitigate this threat, we chose to use a keyword extraction method that has been proven to be effective in previous studies \citep{ren2019neural,li2022identifying}.

Furthermore, we established links among commit messages, pull requests, and issues, through the issue IDs or pull request IDs. These IDs are created according to the Apache development guidelines. The extent to which developers follow the guidelines has an impact on construct validity.
If these guidelines are not correctly followed, we could have missing or wrong links.

A final threat concerns the SATD relation analysis.
Specifically, it is common that two SATD items are related regarding textual information, but actually they refer to different technical debt items.
For example, \textit{fix typo - [from Camel-6b5f64a]} and \textit{typo - [from Camel-a41323c]} both describe typos but they refer to different typos.
To reduce this risk, similarly to previous studies that captured SATD repayment information from linked commit messages \citep{iammarino2019self,zampetti2018self}, we focused on the SATD relation analysis in the same contribution flows.

\subsection{Reliability}

Reliability reflects the extent to which the data and analysis are dependent on the particular researchers.
The first and most important measure we took in mitigating this bias, is to design the study using the well-established case study guidelines proposed by \cite{runeson2012case}. The study design was reviewed by all three authors and iteratively improved during the course of the study.

Furthermore, in this work, we manually classified pull requests and commit messages into different types of SATD or non-SATD.
To reduce the bias caused by manual analysis, after the first author annotated the data, the second author analyzed a part of the sample with the size greater than the statistically significant sample size (i.e., 372). 
Then the disagreements were discussed among the three authors and Cohen's kappa coefficient \citep{landis1977measurement} was calculated. 
The results showed that we have achieved `substantial' agreement \citep{landis1977measurement} with Cohen's kappa coefficient of $+0.74$.
This means more than a simple majority, but not necessarily unanimity.

Moreover, when investigating the relations between SATD in different sources, the first and second authors independently analyzed a sample of 200 pairs of SATD items.
Then Cohen's kappa coefficient was calculated to be $+0.81$, which is considered to be an `almost perfect' agreement \citep{landis1977measurement}.

\subsection{External validity}

This aspect of validity concerns the generalizability of our findings.
Because we utilized supervised machine learning techniques to identify SATD, the generalizability of the training data from different sources has a huge impact on the generalizability of our findings.
Thus, we selected two publicly available SATD datasets in source code comments \citep{da2017using} and issue tracking systems \citep{li2022identifying} as they are gathered from several well-known open-source projects.

Moreover, since there was no dataset available in pull requests and commit messages, we collected and analyzed data from 103 Apache open-source projects. %
Specifically, we manually classified 5,000 pull request sections and 5,000 commit messages because our previous work reported that 3,400 pieces of data are sufficient for a similar SATD identification task \citep{li2022identifying}.

Furthermore, we used stratified 10-fold cross-validation to evaluate the predictive performance of machine learning approaches to reduce bias during training and testing.
However, because all training data is from open-source projects, considering the differences between open-source projects and industrial projects (e.g., differences in technical debt tracking), there are limitations to generalizing results to industry projects.

Overall, our findings can be generalized to other open-source projects of similar size and complexity.
Additionally, different organizations might have different rules for SATD documentation; thus, the extracted keywords might not be generalizable to other projects.

Additionally, similarly to previous approaches \citep{da2017using,ren2019neural,li2022identifying}, as our proposed machine learning approach aims to predict a single type of SATD from the input text, it is not able to identify more than one, if the input text indicates multiple SATD items.

\section{Conclusion}
\label{sec:conclusion}

In this work, we proposed an approach (MT-Text-CNN) to automatically identify four types of SATD (i.e., code/design, documentation, requirement, and test debt) from different sources, namely source code comments, issue tracking systems, pull requests, and commit messages.
Our approach outperformed all baseline methods with an average F1-score of 0.611.
Following that, we summarized and presented lists of SATD keywords.
We found that issues and pull requests are the two most similar sources regarding the number of shared SATD keywords, followed by commit messages, and then followed by code comments.
Thereafter, we applied the MT-Text-CNN approach to characterize SATD in 103 open-source projects and found that SATD is evenly spread among four different sources.
Finally, we explored the relations between SATD in different sources and found that there are four types of relations between SATD in distinct sources.

\bibliographystyle{spbasic}
\bibliography{main}

\end{document}